\documentclass[,
preprint,
superscriptaddress,
amsmath,amssymb,
aps,
prc,
]{revtex4}
\usepackage{mathrsfs}
\usepackage[pdftex]{graphicx}
\usepackage{dcolumn}
\usepackage{bm}
\usepackage{longtable}
\usepackage{color,hyperref}
\usepackage{longtable}
\hypersetup{colorlinks,breaklinks,linkcolor=blue,urlcolor=blue,anchorcolor=blue,citecolor=blue}

\begin{document}
\title{The role of the $\nu g_{9/2}$ orbital in the development of collectivity in the $A\sim60$ region: The case of $^{61}$Co }
\author{A. D. Ayangeakaa}
 \email{ayangeakaa@anl.gov}
\author{S. Zhu}
\author{R. V. F. Janssens}
\author{M. P. Carpenter}
\affiliation{Physics Division, Argonne National Laboratory, Argonne, Illinois 60439, USA}

\author{M. Albers}
\altaffiliation[Present Address: ]{Ernst \& Young GmbH, Mergenthalerallee 3-5, D-65760 Eschborn, Germany.}
\affiliation{Physics Division, Argonne National Laboratory, Argonne, Illinois 60439, USA}

\author{M. Alcorta}
\altaffiliation[Present Address: ]{TRIUMF, 4004 Wesbrook Mall, Vancouver, British Columbia, V6T 2A3 Canada.}
\affiliation{Physics Division, Argonne National Laboratory, Argonne, Illinois 60439, USA}

\author{T. Baugher}
\affiliation{National Superconducting Cyclotron Laboratory, Michigan State University, East Lansing, Michigan 48824, USA}
\affiliation{Department of Physics and Astronomy, Michigan State University, East Lansing, Michigan 48824, USA}

\author{P. F. Bertone}
\altaffiliation[Present Address: ]{Marshall Space Flight Center, Building 4600 Rideout Rd, Huntsville, Alabama 35812, USA.}
\affiliation{Physics Division, Argonne National Laboratory, Argonne, Illinois 60439, USA}

\author{C. J. Chiara}
\altaffiliation[Present Address: ]{U.S. Army Research Laboratory, Adelphi, Maryland 20783, USA.}
\affiliation{Physics Division, Argonne National Laboratory, Argonne, Illinois 60439, USA}
\affiliation{Department of Chemistry and Biochemistry, University of Maryland, College Park, Maryland 20742, USA}

\author{P. Chowdhury} 
\affiliation{Department of Physics, University of Massachusetts, Lowell, Massachusetts 01854, USA}

\author{H. M. David}
\affiliation{Physics Division, Argonne National Laboratory, Argonne, Illinois 60439, USA}

\author{A. N. Deacon}

\affiliation{School of Physics and Astronomy, Schuster Laboratory, University of Manchester, Manchester M13 9PL, United Kingdom}

\author{B. DiGiovine}
\affiliation{Physics Division, Argonne National Laboratory, Argonne, Illinois 60439, USA}

\author{A. Gade}
\affiliation{National Superconducting Cyclotron Laboratory, Michigan State University, East Lansing, Michigan 48824, USA}
\affiliation{Department of Physics and Astronomy, Michigan State University, East Lansing, Michigan 48824, USA}

\author{C. R. Hoffman}
\affiliation{Physics Division, Argonne National Laboratory, Argonne, Illinois 60439, USA}

\author{F. G. Kondev}
\affiliation{Nuclear Engineering Division, Argonne National Laboratory, Argonne, Illinois 60439, USA}

\author{T. Lauritsen}
\affiliation{Physics Division, Argonne National Laboratory, Argonne, Illinois 60439, USA}

\author{C. J. Lister}\altaffiliation[Present Address: ]{Department of Physics, University of Massachusetts, Lowell, Massachusetts 01854, USA.}
\affiliation{Physics Division, Argonne National Laboratory, Argonne, Illinois 60439, USA}

\author{E. A. McCutchan}
\altaffiliation[Present Address: ]{National Nuclear Data Center, Brookhaven National Laboratory, Upton, New York 11973-5000, USA.}
\affiliation{Physics Division, Argonne National Laboratory, Argonne, Illinois 60439, USA}

\author{D. S. Moerland}
\affiliation{Physics Division, Argonne National Laboratory, Argonne, Illinois 60439, USA}
\affiliation{Department of Physics, Florida State University, Tallahassee, Florida 32306, USA}

\author{C. Nair}
\affiliation{Physics Division, Argonne National Laboratory, Argonne, Illinois 60439, USA}

\author{A. M. Rogers} 
\altaffiliation[Present Address: ]{Department of Physics, University of Massachusetts, Lowell, Massachusetts 01854, USA.}
\affiliation{Physics Division, Argonne National Laboratory, Argonne, Illinois 60439, USA}

\author{D. Seweryniak}
\affiliation{Physics Division, Argonne National Laboratory, Argonne, Illinois 60439, USA}

\date{\today}

\begin{abstract}
An extensive study of the level structure of $^{61}$Co has been performed following the complex $^{26}$Mg($^{48}$Ca, \;$2\alpha 4np\gamma$)$^{61}$Co reaction at beam energies of 275, 290 and 320 MeV using Gammasphere and the Fragment Mass Analyzer (FMA). The low-spin structure is discussed within the framework of shell-model calculations using the GXPF1A effective interaction. Two quasi-rotational bands consisting of stretched-$E2$ transitions have been established up to spins $I=41/2$ and $(43/2)$, and excitation energies of $\sim17$ and $\sim20$ MeV, respectively. These are interpreted as signature partners built on a neutron $\nu (g_{9/2})^2$  configuration coupled to a proton $\pi p_{3/2}$ state, based on Cranked Shell Model (CSM) calculations and comparisons with observations in neighboring nuclei. In addition, four $\Delta I=1$ bands were populated to high spin, with the yrast dipole band interpreted as a possible candidate for the shears mechanism, a process seldom observed thus far in this mass region. 
 
\end{abstract}
\maketitle

\section{Introduction}
It is by now well established that the neutron $\nu g_{9/2}$ intruder orbital plays an important role in the development of nuclear structure and the description of high-spin phenomena in neutron-rich nuclei of the $A\sim60$ mass region. For example, prolate-deformed configurations, built upon single-particle excitations, have been observed at moderate to high spins in neutron-rich isotopes of $_{24}$Cr~\cite{PhysRevC.74.064315-zhu,Deacon2005151}, $_{25}$Mn~\cite{PhysRevC.81.014305,PhysRevC.82.054313}, and $_{26}$Fe~\cite{PhysRevC.76.054303,PhysRevC.82.044305}, and interpreted using configurations involving the $\nu g_{9/2}$ orbital~\cite{PRC.87.041305-mpcarpenter}. Moreover, large-scale shell-model calculations performed in the full $fp$ shell provide further corroborating evidence for the need to also include the $g_{9/2}$ orbital in a successful and consistent description~\cite{PhysRevC.82.044305,PhysRevC.82.044316,PhysRevC.85.051301} of these nuclei. The emergence of collective effects in this region, as demonstrated by microscopic mean-field calculations~\cite{PhysRevLett.97.162501-otsuka,Sorlin2008602}, relates directly to the weakening of the attractive monopole part of the tensor interaction between the $\pi f_{7/2}$ and $\nu f_{5/2}$ single-particle orbitals. In the transition from nickel ($Z=28$) to calcium ($Z=20$), which corresponds to the removal of protons from the $\pi f_{7/2}$ orbital, the reduced tensor force generates an upward shift in the energy of the $\nu f_{5/2}$ orbital, which consequently reduces the gap between the $\nu f_{5/2}$ and $\nu g_{9/2}$  single-particle states. The compression of these levels, in turn, allows for the emergence of new subshell closures in exotic nuclei~\cite{Janssens200255,PRC.70.064303-liddick,stepp-nature} and the possibility of pairwise excitations of low-orbit neutrons into the deformation-driving $ g_{9/2}$ orbital, leading to the development of sizable collectivity at medium to high spins in mid-shell nuclei. In fact, a shape coexistence picture appears to emerge at moderate spin, at least in the Cr and Fe isotopic chains~\cite{PRC.87.041305-mpcarpenter}. Moreover, recent data in $^{68,70}$Ni provide first evidence for shape coexistence at low spin in these nuclei~\cite{PhysRevC.89.021301,PhysRevC.88.041302,Chiara70Ni}.

The evolution of nuclear shell structure induced by the weakening of the attractive nucleon-nucleon tensor interaction has been well documented in this region: In the Cr and Fe isotopes, a systematic compilation of the first $2^+$ and $4^+$ states points to a steady decrease in energy as $N$ increases towards $N=40$~\cite{PhysRevC.81.051304,PhysRevC.77.054306}, with $^{64}$Cr exhibiting the lowest $2^+_1$ state among the known $N=40$ isotones~\cite{PhysRevC.81.051304}. The enhancement of collectivity implied by the energy systematics is supported further by intermediate-energy Coulomb excitation and excited-state lifetime measurements~\cite{PhysRevLett.110.242701,PhysRevC.86.011305,PhysRevC.81.061301,PhysRevLett.106.022502}. Furthermore, rotational band structures associated with highly-deformed quadrupole shapes have been observed at high spins in $^{56,57,58,59,60}$Ni~\cite{PhysRevLett.82.3763,Reviol200128,PhysRevLett.86.1450,PhysRevC.65.061302,PhysRevC.78.054318}, and quite recently, in the more neutron-rich isotopes $^{62}$Ni~\cite{ni62-albers} and $^{63}$Ni~\cite{PhysRevC.88.054314-albers} located closer to the $N=40$ shell closure. Identification in the latter two cases was possible due to the implementation of a novel experimental multi-nucleon transfer technique that enabled the production of these nuclei at high spins. 

For the lighter cobalt isotopes, much of the known low-spin structure is well described  
by configurations involving particle-hole excitations among the $p_{3/2}$, $f_{5/2}$, and $p_{1/2}$ single-particle states (see, for example, Refs.~\cite{NSR2006SI37,NSR1975BE22}). The only known case of particle excitations involving the $\nu g_{9/2}$ orbital is in $^{57}$Co~\cite{PhysRevC.65.034309}, where a pair of highly-deformed rotational bands was described as two signature-partner sequences based on a $\nu (g_{9/2})^1$ configuration. Other than this, no experimental evidence for collective excitations involving the $\nu g_{9/2}$ orbital exists for the Co isotopes near $N=40$. In this report, we present results on the observation of high-spin deformed bands in $^{61}$Co, produced via the  high-energy, inverse-kinematics  reaction, $^{26}$Mg($^{48}$Ca, \;$2\alpha 4np\gamma$)$^{61}$Co. The low-spin states are interpreted in the shell-model framework using the GXPF1A effective interaction. The observed high-spin bands are compared with similar structures in neighboring nuclei and with results of calculations within the framework of the cranked shell model (CSM).

\section{Experiment}
Excited states in $^{61}$Co were populated in the multi-nucleon transfer reaction, $^{26}$Mg($^{48}$Ca, \;$2\alpha 4np\gamma$)$^{61}$Co, in inverse kinematics. A self-supporting  0.973-mg/cm$^2$-thick $^{26}$Mg target was bombarded by a series of 275-, 290-, and 320-MeV $^{48}$Ca beams supplied by the Argonne Tandem Linear Accelerator System (ATLAS).  These energies were chosen to be roughly 200\% above the Coulomb barrier in order to favor multi-nucleon transfer processes and, in turn, enhance the population of mostly yrast  and near-yrast states up to fairly high angular momenta. Gamma rays emitted in the de-excitation process were detected with Gammasphere~\cite{I.Y.Lee1990NPA}, a 4$\pi$ array of 101 Compton-suppressed high-purity germanium (HPGe) detectors. The reaction residues were transported to the focal plane of the Fragment Mass Analyzer (FMA), where they were dispersed according to their mass-to-charge ratios, $M/q$. The FMA was tuned for the optimum transport of ions with an average charge state of $19^+$. The recoils were identified on an event-by-event basis from the position and time-of-flight measured in a micro-channel plate detector (MCP) placed at the focal plane and the energy loss measured with a three-fold segmented ionization chamber positioned behind the focal plane. 
The events were accumulated and recorded under the condition that recoil products be detected in coincidence with two or more $\gamma$ rays within a 50-ns time window. Particle identification plots as well as specific details regarding the isotopic selection techniques and the overall experimental procedure can be found in an earlier report on this experiment - see Ref.~\cite{PhysRevC.88.054314-albers}. The accumulated events were sorted into fully-symmetrized two-dimensional $E_{\gamma}$-$E_{\gamma}$ coincidence matrices and analyzed with the \textsc{radware} analysis package~\cite{Radford1995}. 
 \begin{center}
\begin{figure}[t!]
\includegraphics[width=0.45\textwidth]{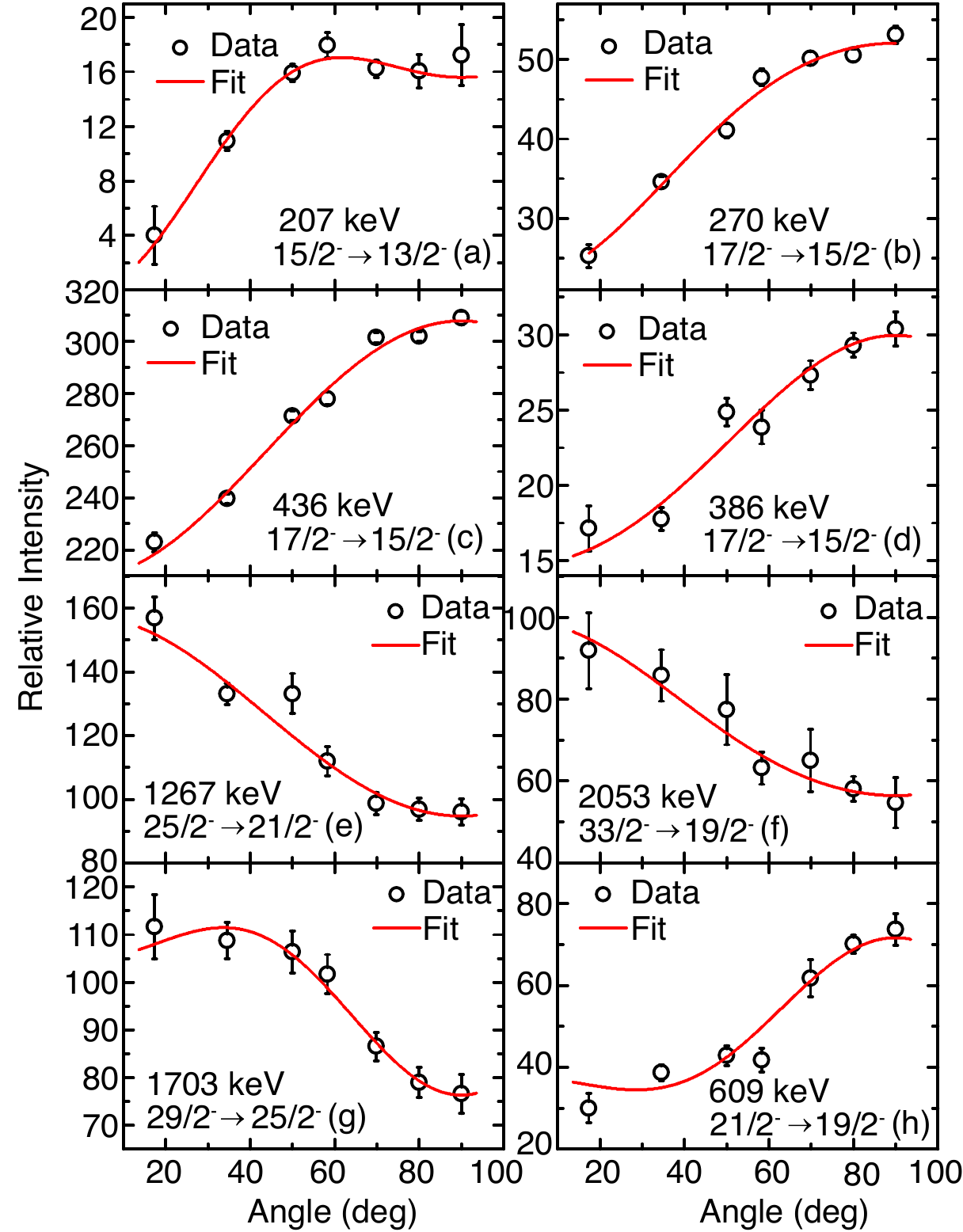}
\caption{\label{fig:angdist} (Color online) Representative angular distributions for newly identified transitions in $^{61}$Co. The solid lines represent least-squares fits using the Legendre polynomial expansion, $W(\theta)=a_{o}[1+a_2P_2(cos\theta) + a_4P_4(cos\theta)]$. Experimental data points are represented by open circles.}
\end{figure}
\end{center}

\begin{figure*}[t!]
\includegraphics[width=0.97\textwidth]{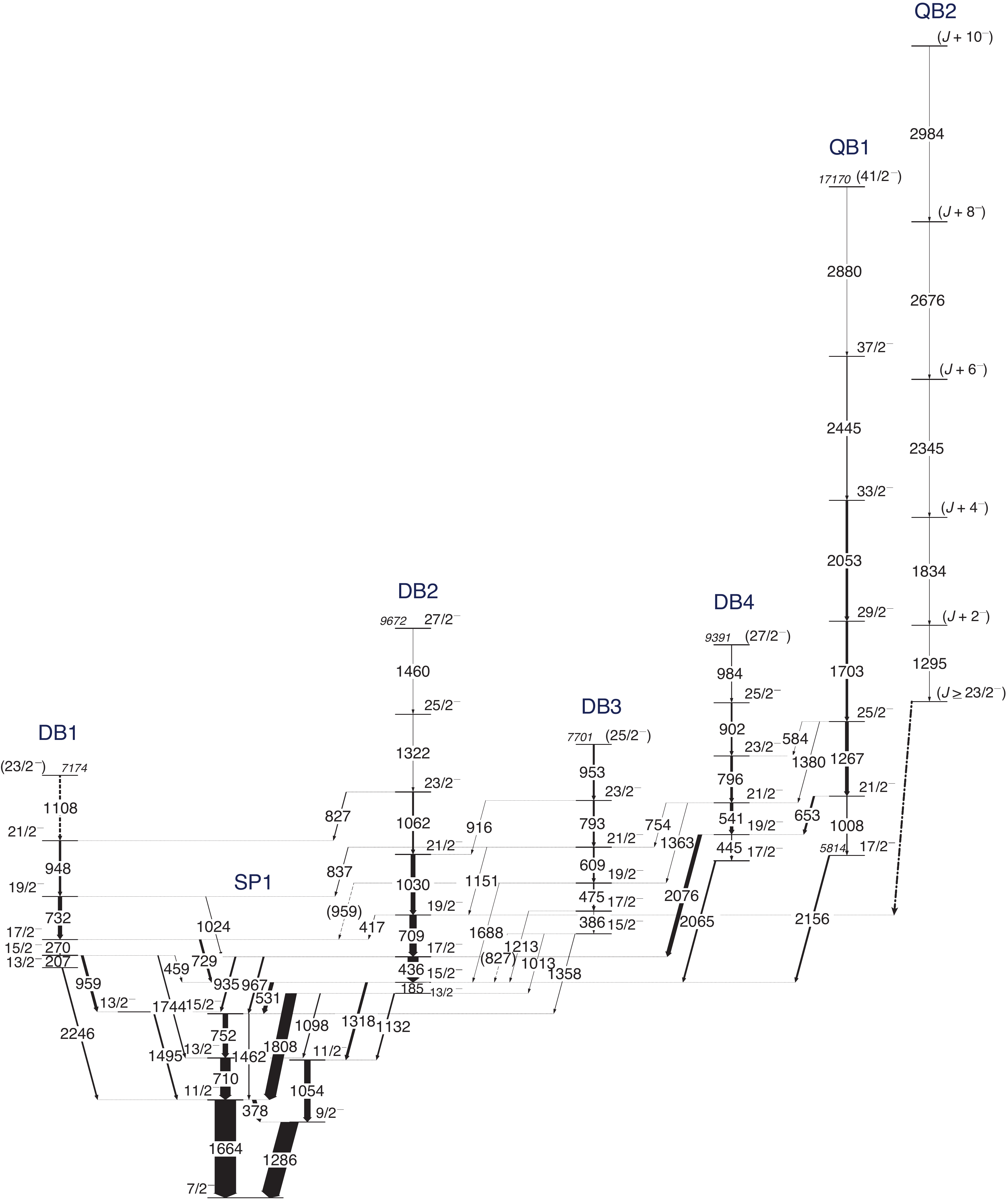}
\caption{\label{fig:levelsch} The level structure of $^{61}$Co  as obtained from the present study. The widths of the arrows are proportional to the relative intensities of the $\gamma$ rays. Tentative transitions are indicated by dashed lines. Note that the excitation energy of band $QB2$ is unknown (marked dash-dotted line) - see text for details.}
\end{figure*}

\vspace{-1cm}
Multipolarities of the newly identified transitions were deduced from the measurements of angular distributions and, for weak ones, from a two-point angular-correlation ratio, $R_{ac}$. The angular-distribution analysis was performed using coincidence matrices sorted in such a way that energies of $\gamma$ rays detected at specific Gammasphere angles (measured with respect to the beam direction) $E_{\gamma}(\theta)$, were incremented on one axis, while the energies of coincident $\gamma$ rays detected at any angle, $E_{\gamma}(any)$, were placed on the other axis. To improve statistics, adjacent rings of Gammasphere and those corresponding to angles symmetric with respect to $90^\circ$ in the forward and backward hemispheres were combined. A total of seven matrices (with the angles 17.3$^\circ$, 34.6$^\circ$, 50.1$^\circ$, 58.3$^\circ$, 69.8$^\circ$, 80.0$^\circ$, and 90.0$^\circ$)~\cite{gamma-doc} were created. After gating on the $E_{\gamma}(any)$ axis, background-subtracted and efficiency-corrected spectra were generated. From these, the intensities of transitions of interest were extracted and fitted to the angular distribution function $W(\theta)=a_{o}[1+a_2P_2(cos\theta) + a_4P_4(cos\theta)]$, where $P_2$ and $P_4$ are Legendre polynomials. The extracted coefficients, $a_2$ and $a_4$, contain the information about the multipolarity of the transitions. Representative fits of angular distributions for some transitions of interest (see below) are displayed in Fig.~\ref{fig:angdist}. 

Transitions for which an angular-distribution analysis was not possible, due to limited statistics, a normalized ratio of $\gamma$-ray intensities observed in detectors in the forward/backward angles to the intensities in detectors centered around $90^\circ$ was determined. For this purpose, two coincident matrices were incremented: In the first, $E_{\gamma}(f/b)$-vs-$E_{\gamma}(any)$, detectors in the forward and backward angles were combined and the matrix incremented such that $\gamma$ rays detected at the 31.7$^\circ$, 37.4$^\circ$, 142.6$^\circ$, 148.3$^\circ$, and 162.7$^\circ$ angles were placed on one axis, with $\gamma$ rays observed at any angle grouped along the other. The second matrix, $E_{\gamma}(\sim90^\circ)$-vs-$E_{\gamma}(any)$, was incremented in a similar fashion, but with transitions observed in detectors at 79.2$^\circ$, 80.7$^\circ$, 90.0$^\circ$, 99.3$^\circ$, and 100.8$^\circ$ degrees placed on one axis. The two-dimensional angular correlation ratio, defined by $R_{ac}$ = $I_{\gamma} (\theta_{f/b}, any)$/$I_{\gamma}(\theta_{ \sim90^\circ},any)$, where $I_{\gamma}(\theta_x, any)$ is the $\gamma$-ray intensity obtained by placing gates on the corresponding $E_{\gamma}(any)$ axis. This ratio, which is independent of the multipolarity of the gating transition, was established to be greater than 1.0 for stretched-quadrupole and less than $0.8$ for stretched-dipole transitions. The energies, relative intensities, and associated angular-distribution coefficients and $R_{ac}$ ratios as well as the multipolarity assignments for the observed transitions are presented in Table~\ref{tab:1}. 

\begin{figure}[t!]
\includegraphics[width=0.45\textwidth]{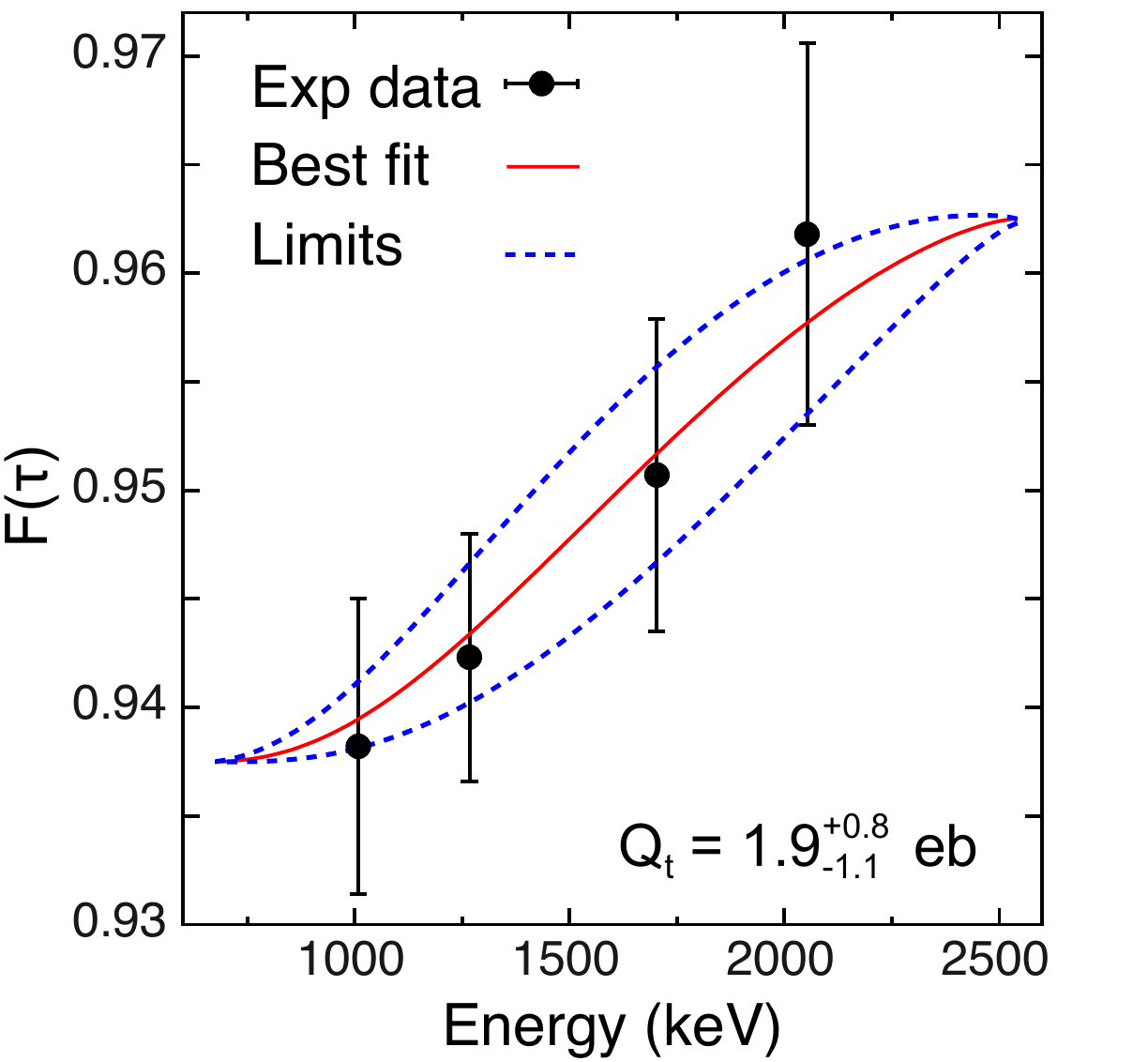}
\caption{\label{fig:ftau} (Color online) Experimental $F(\tau)$ values as a function of the $\gamma$-ray energy (filled circles) compared with the best-fit curve (solid line) for band $QB1$ in $^{61}$Co. The dashed lines indicate the statistical errors only; i.e., they do not include the $\sim$$10 \%$ error associated with the systematic uncertainty in the stopping powers. See text for details.}
\end{figure}

Following a procedure similar to that outlined in Ref.~\cite{PhysRevC.88.054314-albers}, a transition quadrupole moment $Q_t$ for the band labeled $QB1$ hereafter was measured using the Doppler-shift attenuation method (DSAM). The measurement was performed using the $E_{beam} =320$ MeV data, which allowed the extraction of fractional Doppler shifts $F(\tau)$  and the associated errors for the most strongly populated states in the $QB1$ cascade. Transitions from these states were emitted, despite using a thin target, while the recoil ions were slowing down inside the  $^{26}$Mg target. These $\gamma$ rays were corrected with a Doppler factor that corresponds to the initial velocity $\beta_o$ of the recoiling ions. This factor was calculated using the reaction kinematics, and the resultant Doppler-corrected data sorted into seven matrices, with coincidence requirement between $\gamma$ rays detected in one specific angle (corresponding to angles at 17.3$^\circ$, 35.6$^\circ$, 50.1$^\circ$, 58.3$^\circ$, 69.8$^\circ$, 80.0$^\circ$, and 90.0$^\circ$)  on one axis and any angle on the other axis. For each angle, the $\gamma$ ray centroids were observed to be slightly shifted, indicating that they were emitted while the recoil ions were slowing down in the target material. Using this information, the average instantaneous recoil velocity $\beta_t$ for each transition was determined from linear fits of the energy shifts as a function of detector angle $\theta$ and the fractional Doppler shift, $F(\tau)=\beta_t/\beta_o$, deduced. A plot of the extracted $F(\tau)$ values as a function of transition energies is presented in Fig.~\ref{fig:ftau}. The transition quadrupole moment, $Q_t$, was obtained by comparing the experimental $F(\tau)$ values to those computed using the Monte Carlo simulation code \textsc{wlife4}~\cite{PhysRevC.55.R2150-moore}, with the stopping powers provided by the SRIM-2010 package~\cite{Ziegler20101818}. To determine the $Q_t$ value using this method, a few commonly used assumptions~\cite{PhysRevC.88.054314-albers} were made: (i) all levels in the $QB1$ cascade were assumed to have the same $Q_t$; (ii) side-feeding into each level was considered to have the same quadrupole moment, $Q_{sf}$ and to be characterized by the same dynamic moment of inertia as the main band into which it feeds; (iii) a parameter $T_{sf}$, which accounted for a one-step side-feeding delay on top of the band, was set to $T_{sf} = 1$ fs throughout the analysis. A $\chi^2$ minimization with the parameters $Q_t$ and $Q_{sf}$ was performed to the experimental $F(\tau)$ values for band $QB1$. The best fit to the data is indicated by the solid red line in Fig.~\ref{fig:ftau}, while the statistical errors, obtained with a $\chi^2$ value increased by one, is represented by the dashed blue lines. An additional $\sim10\%$ systematic error was added to the final result to take into account the uncertainties associated with the simulation of the stopping process. 

\section{level scheme}\label{levelsch}
The level scheme of $^{61}$Co deduced in the present investigation is presented in Fig.~\ref{fig:levelsch}, and the assigned transitions and their properties are summarized in Table~\ref{tab:1}. Two quasi-rotational band structures consisting of stretched-$E2$ transitions were identified and assigned to $^{61}$Co based on gating on the appropriate focal plane information and on observed coincidences with previously known low-lying transitions~\cite{PhysRevC.54.1084-regan,PRC.85.064305-Recchia}. In addition, four $\Delta I = 1$ bands were also identified, along with a number of other levels with single-particle character. As noted earlier, multipolarity assignments are proposed based on the analysis of angular distributions and angular-correlation ratios. In some instances, no definitive parity assignment could be made, since neither linear polarization nor internal conversion was measured. 

\begin{center}
\begin{figure*}[t!]
\includegraphics[width=0.8\textwidth]{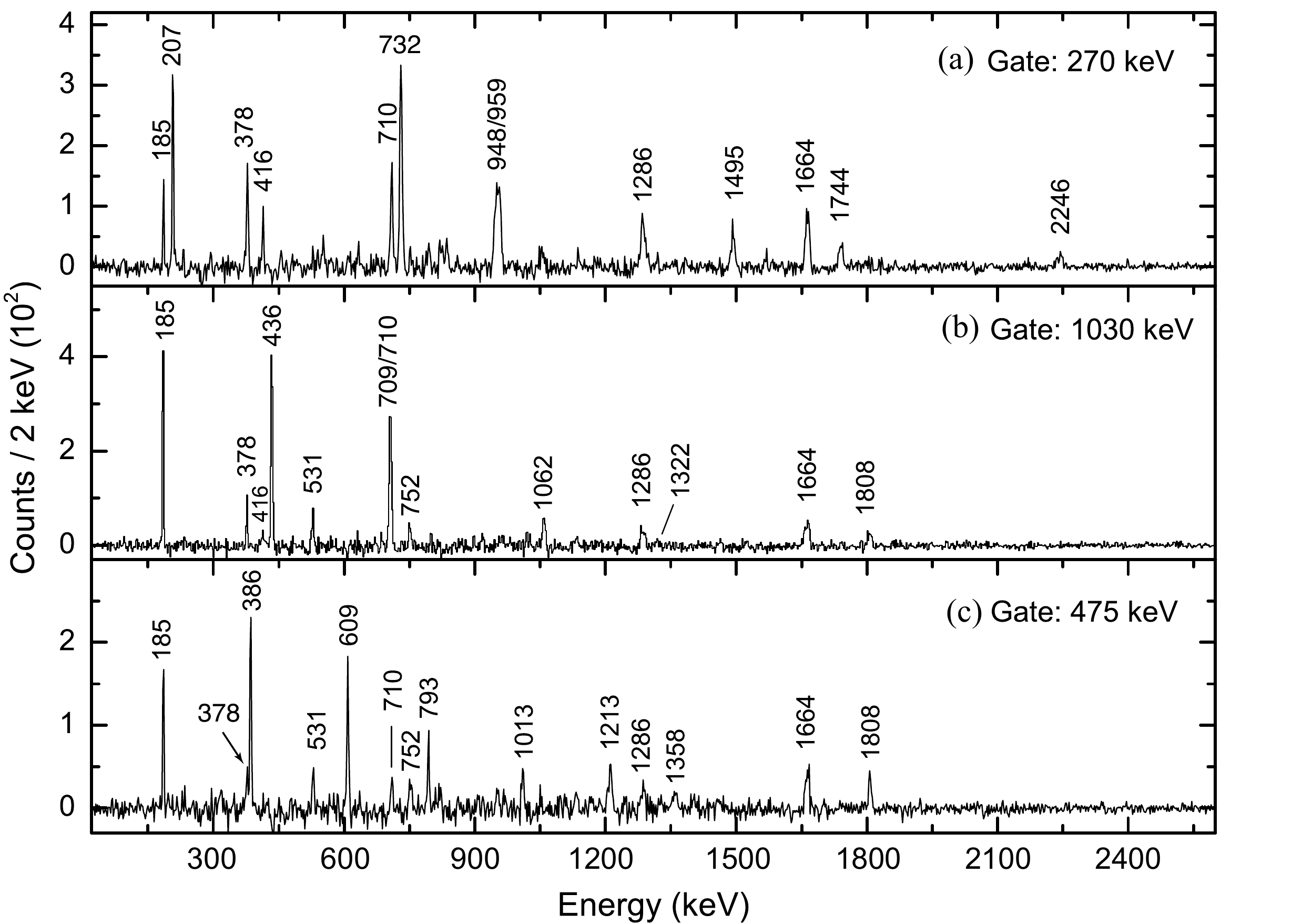}
\caption{\label{fig:sp1} Representative Doppler-corrected coincidence spectra gated by transitions in $^{61}$Co. (a) Gate on the 270-keV transition in the dipole band, $DB1$. (b) Gate on the 1030-keV transition in band $DB2$. (c) Gate on the 475-keV $\gamma$ ray in band $DB3$. Some of the relevant coincidence relationships are highlighted in the text.}
\end{figure*}
\end{center}

\begin{center}
\begin{figure*}[t!]
\includegraphics[width=0.8\textwidth]{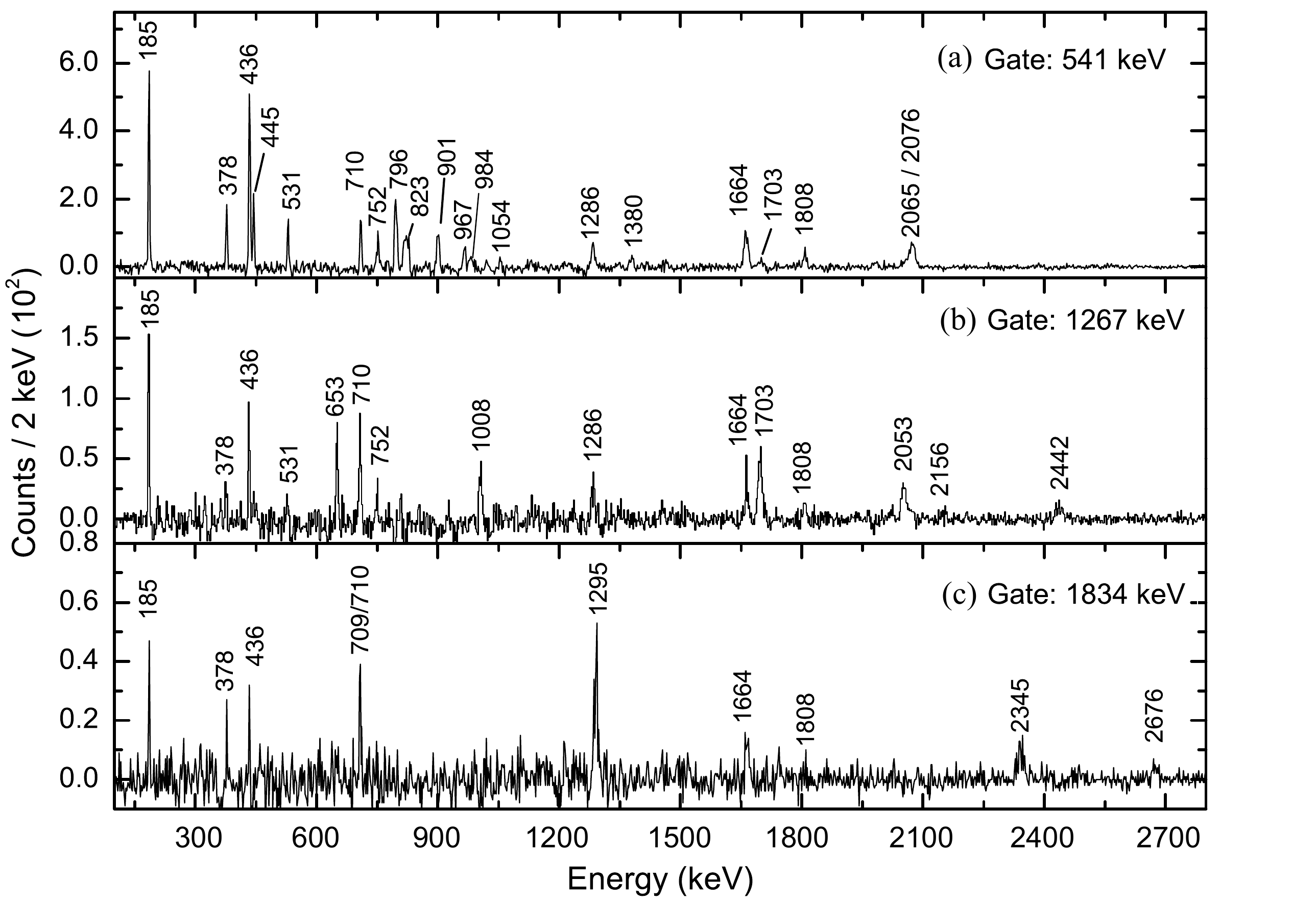}
\caption{\label{fig:sp2} Representative Doppler-corrected coincidence spectra gated by transitions in $^{61}$Co. (a) Gate on the 541-keV transition in the dipole band, $DB4$. (b) Gate on the 1267-keV line in the quadrupole band $QB1$. (c) Gate on the 1834-keV $\gamma$ ray in the quadrupole band $QB2$.}
\end{figure*}
\end{center}

The $^{61}$Co nucleus, with $Z=27$ and $N=34$, has a ground-state spin and parity of $7/2^-$ due to the presence of a proton hole in the $f_{7/2}$ single-particle state. This $I^\pi$ value has been confirmed experimentally from the $\beta$ decay of $^{61}$Fe~\cite{PRC.11.966-bron}. The low-lying levels built on the ground state similarly have negative parity and are understood as being due to the occupation of the $f$$p$ neutron orbitals near the Fermi surface and the coupling with the $f_{7/2}$ proton hole. These levels, which are grouped together and identified as $SP1$ in Fig.~\ref{fig:levelsch}, have been reported previously in the works of Regan {\it et al.}~\cite{PhysRevC.54.1084-regan} and Recchia {\it et al.}~\cite{PRC.85.064305-Recchia}. The latter two studies represent the most recent investigations of $^{61}$Co, in which excited states were populated up to the $19/2^-$ level at 4803 keV. The placement of these levels in the decay scheme is confirmed here, the only exception being the 298- and 1028-keV transitions (reported in Ref.~\cite{PhysRevC.54.1084-regan}). These two $\gamma$ rays were not observed in this investigation, although a 1030-keV line was identified, but assigned as a member of the $DB2$ cascade based on coincidence relationships (more details below). The remaining band-like structures, labeled $DB1$, $DB3$, and $DB4$ for `dipole bands', and $QB1$, and $QB2$ for `quadrupole bands' are essentially new to this work and are the main focus of the present investigation. Representative $\gamma$-ray spectra, obtained by placing single coincidence gates on $\gamma$ rays in the new structures, are presented in Figs.~\ref{fig:sp1} and \ref{fig:sp2}.

Figure~\ref{fig:sp1}(a) results from a coincidence gate on the 270-keV transition in the $DB1$ cascade. The band consists of a regular sequence of $\Delta I =1$ transitions that extends, tentatively, up to $I^\pi$ = $(23/2^-)$ at 7174 keV. The multipole character of the in-band transitions was deduced from the measured angular-distribution coefficients and the correlation ratios. Typical angular distribution plots for $\Delta I=1$ transitions are presented in Figs.~\ref{fig:angdist}(a) and (b),  for the 207- and 270-keV transitions, respectively. Due to limited statistics, it was not possible to perform a multipolarity analysis for the 1108-keV transition, and a dipole character was assumed. Furthermore, two unresolved doublets relating to this band were observed: The 948-/959- [see Fig~\ref{fig:sp1}(a)], and the 729-/732-keV doublets. While it was not possible to differentiate these doublets in the present multipolarity analysis, the summed angular distributions for both were observed to be consistent with an $M1/E2$ mixed character and, hence, assumed to be dipole in nature. The linking of this sequence to the $SP1$ structure was facilitated by the observation of the quartet of 959-, 1495-, 1744-, and 2246-keV transitions. Similar to the in-band transitions, the linking transitions are characterized by an $M1/E2$ admixture, as deduced from the measured angular-correlation ratios and/or angular-distribution coefficients. These observations firmly establish the $13/2^-$ bandhead of the $DB1$ sequence. In addition, two weak $\gamma$ rays of 552 and 636 keV were observed in coincidence with this band, but could not be unambiguously placed in the level scheme.

\vspace{-2.5cm}
The $DB2$ sequence in Fig.~\ref{fig:levelsch} is the most intense band built on the $SP1$ single-particle structure, and consists mainly of $\Delta I = 1$ transitions (see Fig.~\ref{fig:angdist}(c), for an example) with no $E2$ crossovers. Figure~\ref{fig:sp1}(b) shows a coincidence spectrum obtained with a gate on the 1030-keV transition depopulating the $21/2^-$ level at 5832 keV. The transitions from this level and the states above it are in coincidence with the previously known 185-, 436-, and 709-keV $\gamma$ rays and have been grouped as a band, herewith extending the $DB2$ band up to a $27/2^-$ state at 9672 keV. The ordering of the transitions within the cascade was supported by the observed decreasing intensities and reinforced by the presence of the 827-keV $\gamma$ ray linking the $23/2^-$, 6893-keV level in band $DB2$ with the  $21/2^-$ state at 6066 keV in $DB1$. The assigned $\Delta I =1$ character of the in-band transitions is based on the measured $a_2$, $a_4$ coefficients and the $R_{ac}$ ratios, as given in Table~\ref{tab:1}. As previously noted~\cite{PhysRevC.54.1084-regan}, this band decays primarily to the lower-lying structure via the 1808- and 1132-keV transitions. The deduced  $a_2$ values of $-0.31(4)$ and $-0.11(2)$ for the 1808- and 1132-keV lines, respectively, are consistent with dipole radiation. 

Much like $DB2$, bands $DB3$ and $DB4$ also consist of sequences with $\Delta I=1$ transitions. Band $DB3$ is built on the $15/2^-$ state at 4485 keV and extends up to the $(25/2^-)$ level at 7701 keV. It is composed of the 386-, 475-, 609-, 793-, and 953-keV transitions, and decays into, and couples very strongly with, band $DB2$. Except for the 1688-keV transition whose $R_{ac}$ ratio of 1.74(5) favors a stretched quadrupole character, the links are predominately dipole in nature. This implies that bands $DB2$ and $DB3$ have the same parity. A coincidence gate on the 475-keV transition, presented in Fig~\ref{fig:sp1}(c), displays the in-band 386-, 609-, 793-keV dipole $\gamma$ rays, and the 1013- and 1213-keV linking transitions. Again, the dipole character of the in-band transitions was deduced from the measured angular-distribution coefficients and $R_{ac}$ ratios. A sample plot of the angular distribution for the 386-keV $\gamma$ ray is presented in Fig.~\ref{fig:angdist}(d). It was not possible to distinguish between the 793- and 796-keV (in band $DB4$) doublet in the present multipolarity analysis, but the angular-distribution coefficients and $R_{ac}$ ratio for the summed peak were found to be consistent with a magnetic dipole-type transition and, hence, the $\Delta I = 1$ assignment. Furthermore, due to limited statistics, no multipolarity measurement was performed for the 953-keV line. The assigned $\Delta I = 1$ character was based on the simple assumption of the continuation of the sequence with transitions of the same character. Similarly, band $DB4$, built on the $17/2^-$ level at 5724 keV decays predominantly into $DB2$ via the 2065- and 2076-keV dipole transitions - see Fig~\ref{fig:sp2}(a). This spectrum, obtained with a coincidence gate on the 541-keV $\gamma$ ray, also indicates the presence of a sizable line at 823 keV that could not be placed in the present level scheme. A gate on this transition appears to be in coincidence with members of bands $DB4$ and $DB1$, but also with the 332-, 690-, and 884-keV transitions in $^{60}$Co~\cite{Taylor197877}. As a result, this line is left out of the current discussion.

Two bands, labeled $QB1$ and $QB2$ in Fig.~\ref{fig:levelsch}, were populated to higher spins and excitation energies: These cascades, which have never been observed before, consist of regular sequences of $\Delta I =2$ transitions.  A coincidence gate on the 1267-keV $\gamma$ ray in band $QB1$ is presented in Fig.~\ref{fig:sp2}(b). The 1008-, 1267-, 1703-, 2053-, 2442-, and 2880-keV cascade constituting band $QB1$ is built on the $17/2^-$, 5814-keV state, and extends up to a tentative spin and parity  of ($41/2^-$) at 17170 keV. It decays predominately into band $DB4$ via the 653-keV transition from the $21/2^-$, 6821-keV level and, very weakly, through the 2156-keV line into the $DB2$ sequence. As presented in Fig.~\ref{fig:angdist}, the angular distributions of the in-band 1267-, 1703- and 2053-keV transitions are all indicative of a stretched-quadrupole nature. Similarly, the angular-distribution coefficients of $a_2 = -0.33(3)$ and $a_4=-0.13(9)$ are consistent with the dipole character assigned to the 653-keV $\gamma$ ray. This, together with the deduced dipole nature of the 2076-keV transition (band $DB4$) and the known multipolarities of the 185- and 436-keV $\gamma$ rays (band $DB2$) fixes the spin and parity of the 6821-keV level as $21/2^-$ (band $QB1$). Using the spin of this level and the deduced dipole nature of the 2156-keV $\gamma$ ray [$R_{ac}=0.87(5)$], a quadrupole character was assigned to the 1008-keV transition. This assignment is proposed in spite of the fact that this 1008-keV $\gamma$ ray forms a doublet with the 1013-keV transition linking the $15/2^-$ state in $DB3$ with the $13/2^-$ level of $DB2$, and that a fit of the angular distribution for the doublet was found to be consistent with a mixed $M1/E2$ multipolarity. Furthermore, the poor statistics at higher energies prevented a firm spin assignment for the 17170-keV level depopulated by the 2880-keV line. The tentative ($41/2^-$) assignment, corresponding to an $E2$ transition, follows from the argument presented earlier of an extension of the sequence by a transition of the same multipolarity. As discussed in the previous section, the centroid-shift Doppler-attenuation method was used in determining the transition quadrupole moment for this band. The extracted experimental $F(\tau)$ values were compared with those simulated with the \textsc{wlife4} code, and a transition quadrupole moment of $Q_t=1.9^{+0.8}_{-1.1}$ eb, corresponding to a quadrupole deformation of $|\beta_2|=0.4(2)$, was obtained.

A second series of quadrupole transitions, labeled $QB2$ in Fig.~\ref{fig:levelsch}, was observed in parallel with band $QB1$. It is populated with the same intensity pattern, but with a lower relative yield. This band is assigned to $^{61}$Co based on the FMA focal plane information and the observed coincidence relationships with lower-lying states, as indicated by the spectrum in Fig.~\ref{fig:sp2}(c). While it has not been possible to determine a definite decay path for the band, the spectrum in Fig.~\ref{fig:sp2}(c), obtained with a coincidence gate on the 1834-keV transition, suggests a feeding to lower-lying states through the $19/2^-$, 4803-keV level in $DB2$. This feeding pattern is also supported by a coincidence gate (not shown) on the 1295-keV transition populating the bandhead of the $QB2$ sequence.  To place this band in the level scheme, several factors were taken into account. For instance, the assumption that the 1295-keV transition might be the link to the $DB$2 band would imply that $QB2$ is yrast relative to $QB1$. This contradicts the experimental observation that band $QB1$ is the most intense of the two quadrupole cascades. Therefore, it was assumed that  the excitation energy of the  bandhead for $QB2$ lies several keV above the $21/2^-$ state in $QB1$ (assuming a quadrupole linking transition). Based on these considerations, and others associated with the interpretation presented in Section~\ref{qbands}, a tentative excitation energy and spin-parity greater or equal to 8.4 MeV and $23/2^-$ were assigned, as illustrated in Fig.~\ref{fig:levelsch}. The proposed negative parity is due to the purported linking between bands $QB2$ and $DB2$. Unlike band $QB1$, no transition quadrupole moment was extracted for this sequence due to the weaker intensity.

\renewcommand{\thefootnote}{\alph{footnote}}
\begin{center}
\LTcapwidth=\textwidth 
\begin{longtable*}{@{\extracolsep{\fill}} rrrrrrcc @{\extracolsep{\fill}}}
\caption{\label{tab:1}Transition energies $E_{\gamma}$, relative intensities $I_{\gamma}$, angular distribution and correlation information for all transitions in $^{61}$Co. The intensities are corrected for detector efficiency and normalized to the 1664.2(4)-keV transition.  $R_{ac}$ is the normalized ratio of $\gamma$-ray intensities in the detectors at forward/backward angles to the intensities in the detectors at angles centered around 90$^{\circ}$. The spin and excitation energy of band $QB2$ are based on $x \ge 8.4 $ MeV and $ J^\pi \ge 23/2^-$ as discussed in the text. Values given in parentheses are tentative.} \\

\hline \hline
\multicolumn{1}{c}{$E_{\gamma}$ (keV)} & \multicolumn{1}{c}{$ I_{\gamma} $} & \multicolumn{1}{c}{$ E_i $ (keV)}  & \multicolumn{1}{c}{$ I^{\pi}_i  \rightarrow  I^{\pi}_f$} &  \multicolumn{1}{c}{$a_2$} & \multicolumn{1}{c}{$a_4$} & \multicolumn{1}{c}{ $R_{ac}$} & \multicolumn{1}{c}{Mult.}\\ \hline 
\endfirsthead

\hline \hline
\multicolumn{1}{c}{$E_{\gamma}$ (keV)} & \multicolumn{1}{c}{$ I_{\gamma} $} & \multicolumn{1}{c}{$ E_i $ (keV)}  & \multicolumn{1}{c}{$ I^{\pi}_i  \rightarrow I^{\pi}_f$} &  \multicolumn{1}{c}{$a_2$} & \multicolumn{1}{c}{$a_4$} & \multicolumn{1}{c}{ $R_{ac}$} & \multicolumn{1}{c}{Mult.}\\ \hline 
\endhead
\hline\hline

\endfoot

\hline \hline
\endlastfoot
 185.1(5)       	& 56.2(1)     	&$    3657.5(2) $&$   15/2^{-} \rightarrow   13/2^{-} $ 	&-0.29(3)	&-0.11(5)&0.72(4)& $M1/E2$ \\
 207.4(8)        	& 3.0(4)       	&$    4116.5(4) $&$   15/2^{-} \rightarrow   13/2^{-} $ 	&-0.56(5)	&-0.40(5)&0.57(2)& $M1/E2$ \\
 269.8(1)      	& 9.9(6)       	&$    4385.9(3) $&$   17/2^{-} \rightarrow   15/2^{-} $ 	&-0.41(5)	&-0.17(6)&0.68(6)& $M1/E2$ \\
 377.8(1)      	& 23.4(8)     	&$    1664.2(4) $&$   11/2^{-} \rightarrow   \;\; 9/2^{-}$ &-0.41(3)	&0.004(3)&0.77(1)& $M1/E2$ \\
 385.8(5)      	& 2.1(4)       	&$    4870.5(5) $&$   17/2^{-} \rightarrow   15/2^{-} $ 	&-0.40(5)	&0.06(6)&0.66(4)& $M1/E2$ \\
 416.5(2)      	& 2.3(1)     	&$    4802.9(4) $&$   19/2^{-} \rightarrow   17/2^{-} $ 	&	&&0.98(7)& $M1/E2$ \\
 435.8(7)       	& 61.5(2)     	&$    4093.2(3) $&$   17/2^{-} \rightarrow   15/2^{-} $ 	&-0.23(1)	&0.004(6)&0.79(7)& $M1/E2$ \\
 445.4(3)      	& 5.1(4)       	&$    6168.7(4) $&$   19/2^{-} \rightarrow   17/2^{-} $ 	&-0.36(1)	&-0.08(3)&0.76(2)& $M1/E2$ \\
 459.0(1)       	& 2.0(2)      	&$    4116.5(4) $&$   15/2^{-} \rightarrow   15/2^{-} $ 	&	&&0.75(2)& $M1/E2$ \\
 475.3(3)        	& 7.3(5)       	&$    5345.5(4) $&$   19/2^{-} \rightarrow   17/2^{-} $ 	&-0.19(2)	&-0.12(3)&1.10(3)& $M1/E2$ \\
 530.6(1)      	& 21.7(8)      	&$    3657.5(2) $&$   15/2^{-} \rightarrow   15/2^{-} $ 	&-0.21(5)	&-0.20(6)&1.35(2)& $M1/E2$ \\
 540.5(2)      	& 19.4(7)      	&$    6708.8(4) $&$   21/2^{-} \rightarrow   19/2^{-} $ 	&-0.19(6)	&-0.07(6)&0.98(4)	& $M1/E2$ \\
  584.1(1)     	 & 1.1(2)     	&$    8088.5(5) $&$   25/2^{-} \rightarrow   23/2^{-} $	&	&	&0.87(5)	&	 $M1/E2$ \\
 608.8(3)        	& 9.2(9)       	&$    5954.5(4) $&$   21/2^{-} \rightarrow   19/2^{-} $ 	&-0.61(8)	&0.28(1)&0.88(3)&	 $M1/E2$ \\
 653.0(2)      	& 9.0(8)       	&$    6821.3(4) $&$   21/2^{-} \rightarrow   19/2^{-} $ 	&-0.33(3)	&-0.13(9)	&0.85(2)	&	 $M1/E2$ \\
 708.6(2)      	& 40.2(2)    	&$    4802.9(4) $&$   19/2^{-} \rightarrow   17/2^{-} $ 	&-0.34(1)	&0.02(2)	&0.78(4)	&	 $M1/E2$ \\
 709.7(1)      	& 58.0(2)     	&$    2374.1(3) $&$   13/2^{-} \rightarrow   11/2^{-} $ 	&-0.31(2)	&-0.05(2)	&0.89(2)	&	 $M1/E2$ \\
 728.5(2)      	& 13.0(6)      	&$    4385.9(3) $&$   17/2^{-} \rightarrow   15/2^{-} $ 	&-0.36(7)	&-0.17(9)	&0.83(3)	&	 $M1/E2$ \\
 731.7(1)      	& 23.0(9)      	&$    5117.6(4) $&$   19/2^{-} \rightarrow   17/2^{-} $ 	&-0.36(7)	&-0.17(9)	&0.83(3)	&	 $M1/E2$ \\
 752.3(1)      	& 30.4(9)      	&$    3126.5(3) $&$   15/2^{-} \rightarrow   13/2^{-} $ 	&-0.18(3)	&-0.03(2)	&0.98(1)	&	 $M1/E2$ \\
 753.8(3)      	& 1.2(3)     	&$    6708.8(4) $&$   21/2^{-} \rightarrow   21/2^{-} $ 	&	&	&0.79(2)	&	 $M1/E2$ \\
 793.0(2)      	& 5.6(2)     	&$    6748.2(4) $&$   23/2^{-} \rightarrow   21/2^{-} $ 	&-0.23(6)	&-0.09(4)	&0.64(7)	&	 $M1/E2$ \\
 795.8(2)      	& 13.7(7)      	&$    7504.5(5) $&$   23/2^{-} \rightarrow   21/2^{-} $ 	&-0.23(6)	&-0.09(4)	&0.64(7)	&	 $M1/E2$ \\
 826.8(3)      	& 4.5(1)     	&$    6892.9(6) $&$   23/2^{-} \rightarrow   21/2^{-} $ 	&			&	&0.86(3)	&	 $M1/E2$ \\	
 (827.3(4))        	& 1.0(2)     	&$    4484.7(5) $&$   15/2^{-} \rightarrow   15/2^{-} $ 	&	&	&0.82(2)	&	 $M1/E2$ \\
 837.3(2)      	& 3.2(3)     	&$    5954.5(4) $&$   21/2^{-} \rightarrow   19/2^{-} $ 	&-0.21(1)	&-0.02(1)	&0.69(3)	&	 $M1/E2$ \\
 901.5(2)      	& 9.2(6)       	&$    8406.8(5) $&$   25/2^{-} \rightarrow   23/2^{-} $ 	&	&	&0.98(6)	&	 $M1/E2$ \\
 915.5(2)		&1.2(3)		&$    6748.2(4) $&$   23/2^{-} \rightarrow   21/2^{-} $ 	&	&	&0.87(2)	&	$M1/E2$ \\	
 935.0(5)       	& 3.0(1)     	&$    4093.2(3) $&$   17/2^{-} \rightarrow   13/2^{-} $ 	&	&	&1.23(2)	&	 $E2$ \\
 947.7(2)      	& 12.0(9)      	&$    6065.9(6) $&$   21/2^{-} \rightarrow   19/2^{-} $ 	&-0.31(3)	&\;0.05(4)	&0.82(3)	&	 $M1/E2$ \\
 952.6(3)        	& 2.2(1)     	&$    7700.9(8) $&$   (25/2^{-}) \rightarrow   23/2^{-} $ 	&	&	&	&	 $M1/E2$ \\
 959.1(3)        	& 15.0(7)      	&$    4116.5(4) $&$   15/2^{-} \rightarrow   13/2^{-} $ 	&-0.31(3)	&\;0.05(4)	&0.82(3)	&	 $M1/E2$ \\
 (959.2(3))        	& 1.1(2)     	&$    5345.5(4) $&$   19/2^{-} \rightarrow   17/2^{-} $ 	&			&	&0.78(2)	&	 $M1/E2$ \\
 967.3(3)        	& 2.1(1)     	&$    4093.2(3) $&$   17/2^{-} \rightarrow   15/2^{-} $ 	&			&	&0.65(3)	&	 $M1/E2$ \\
 983.6(2)     	& 2.5(4)       	&$   9391.2(8) $&$   (27/2^{-}) \rightarrow   25/2^{-} $ 	&			&	& 	&	 $M1/E2$ \\
 1008.1(2)     	& 2.3(9)     	&$    6821.3(4) $&$   21/2^{-} \rightarrow   17/2^{-} $ 	&			&	&1.26(2)	&	 $E2$ \\
 1013.3(6)       	& 1.2(1)     	&$    4484.7(5) $&$   15/2^{-} \rightarrow   13/2^{-} $ 	&-0.41(2)	&-0.21(1)	&0.97(4)	&	 $M1/E2$ \\
 1023.8(1)      	& 3.0(2)      	&$    5117.6(4) $&$   19/2^{-} \rightarrow   17/2^{-} $ 	&			&	&0.96(5)	&	 $M1/E2$ \\
 1030.0(2)     	& 26.3(2)     	&$    5832.0(5) $&$   21/2^{-} \rightarrow   19/2^{-} $ 	&-0.25(1)	&-0.08(1)	&0.87(5)	&	 $M1/E2$ \\
 1053.7(2)     	& 35.5(1)     	&$    2339.7(3) $&$   11/2^{-} \rightarrow    \; \; 9/2^{-} $ &			&	&0.84(2)	&	 $M1/E2$ \\
 1061.8(3)       	& 15.6(5)     	&$    6892.9(6) $&$   23/2^{-} \rightarrow   21/2^{-} $ 	&-0.32(2)	&-0.02(1)	&1.01(5)	&	 $M1/E2$ \\
 1097.6(6)       	& 5.7(5)       	&$    3472.3(2) $&$   13/2^{-} \rightarrow   13/2^{-} $ 	&			&	&0.78(3)	&	 $M1/E2$ \\
 (1108.3(3))         	& 1.0(10)      	&$    7173.9(6) $&$   (23/2^{-}) \rightarrow   21/2^{-} $ 	&			&	&0.92(5)	&	 $M1/E2$ \\
 1132.3(4)       	& 8.5(3)         	&$    3472.3(2) $&$   13/2^{-} \rightarrow   11/2^{-} $ 	&-0.11(2)			&0.38(3)	&0.89(2)	&	 $M1/E2$ \\
 1150.9(3)       	& 1.3(1)    		&$    5954.5(4) $&$   21/2^{-} \rightarrow   19/2^{-} $ 	&-0.56(2)	&0.12(1)	&0.67(3)	&	 $M1/E2$ \\
 1212.8(4)       	& 1.1(2)     	&$    4870.5(5) $&$   17/2^{-} \rightarrow   15/2^{-} $ 	&	&	&0.79(2)	&	 $M1/E2$ \\
 1267.2(1)     	& 18.9(6)      	&$    8088.5(5) $&$   25/2^{-} \rightarrow   21/2^{-} $ 	&0.38(5)	&-0.02(6)&1.51(2)	& $E2$ \\
 1285.9(1)     	& 98.0(2)     	&$    1286.1(2) $&$    9/2^{-} \rightarrow    \; \; 7/2^{-} $ &-0.19(3)	&0.02(4)&1.05(8)& $M1/E2$ \\
 1294.9(1)     	& 2.1(2)    		&$    x + 1295.0(1) $&$ (J + 2^-) \rightarrow J $ 	&0.41(3)	&-0.21(5)	&1.23(2)	& $E2$ \\
 1318.0(3)       	& 14.2(6)      	&$    3657.5(2) $&$   15/2^{-} \rightarrow   11/2^{-} $ &0.51(9)		&-0.16(1)	&1.16(3)	&	 $E2$ \\
 1321.9(2)      	& 2.1(2)     	&$    8212.3(4) $&$   25/2^{-} \rightarrow   23/2^{-} $ &-0.33(7)		&-0.13(1)	&0.94(3)	&	 $M1/E2$ \\
 1358.0(4)       	& 2.3(1)     	&$    4484.7(5) $&$   15/2^{-} \rightarrow   15/2^{-} $ &-0.17(5)		&-0.04(7)	&0.76(4)	&	 $M1/E2$ \\
 1363.4(4)       	& 1.3(3)     	&$    6708.8(4) $&$   21/2^{-} \rightarrow   19/2^{-} $ &		&	&0.97(3)	&	 $M1/E2$ \\
 1379.7(5)       	& 2.1(3)     	&$    8088.5(5) $&$   25/2^{-} \rightarrow   21/2^{-} $ 	&	&	&1.13(3)	&	 $E2$ \\
 1460.1(2)     	& 1.1(1)     	&$    9672.3(4) $&$   27/2^{-} \rightarrow   25/2^{-} $ &			&		&0.87(4)	&	 $M1/E2$ \\
 1462.3(2)      	& 5.0(2)      	&$    3126.5(3) $&$   15/2^{-} \rightarrow   11/2^{-} $ &0.25(3)		&-0.41(3)	&1.22(2)	&	 $E2$ \\
 1495.1(1)      	& 9.2(7)       	&$    3157.5(5) $&$   13/2^{-} \rightarrow   11/2^{-} $ &		&	&0.91(3)	&	 $M1/E2$ \\
 1664.2(4)      	& 120.0(2)    	&$    1664.2(4) $&$   11/2^{-} \rightarrow    \;\;7/2^{-} $ &0.45(3)	&-0.21(3)	&1.13(7)	&	 $E2$ \\
 1687.7(6)       	& 1.0(1)     	&$    5345.5(4) $&$   19/2^{-} \rightarrow   15/2^{-} $ &			&	&1.74(5)	&	 $E2$ \\
 1702.9(2)     	& 13.6(10)     	&$    9791.8(5) $&$   29/2^{-} \rightarrow   25/2^{-} $ &0.28(2)		&-0.16(3)	&1.39(3)	&	 $E2$ \\
 1743.9(8)       	& 6.5(5)       	&$    4116.5(4) $&$   15/2^{-} \rightarrow   13/2^{-} $ &		&	&0.79(4)	&	$M1/E2$ \\
 1808.1(1)     	& 62.9(2)     	&$    3472.3(2) $&$   13/2^{-} \rightarrow   11/2^{-} $ &-0.31(4)		&-0.21(5)	&0.82(7)	&	 $M1/E2$ \\
 1834.1(2)    	& 1.9(2)     	&$  x + 3129.1(7)    $&$ (J + 4^-) \rightarrow (J + 2^-) $ &0.41(6)	&-0.19(7)	&1.63(3)	&	 $E2$ \\
 2052.8(2)     	& 13.7(9)      	&$   11844.9(7) $&$   33/2^{-} \rightarrow   29/2^{-} $ &0.33(5)		&0.02(7)	&1.31(5)	&	 $E2$ \\
 2064.5(6)       	& 9.3(7)       	&$    5723.7(6) $&$   17/2^{-} \rightarrow   15/2^{-} $ &-0.41(3)		&-0.09(1)	&0.78(4)	&	 $M1/E2$ \\
 2076.3(3)       	& 15.2(3)      	&$    6168.7(4) $&$   19/2^{-} \rightarrow   17/2^{-} $ &-0.29(5)		&-0.12(3)	&0.83(2)	&	 $M1/E2$ \\
 2156.2(3)       	& 9.4(5)     	&$    5813.5(5) $&$   17/2^{-} \rightarrow   15/2^{-} $ &		&	&0.87(5)	&	 $M1/E2$ \\
 2246.0(1)      	& 8.2(9)       	&$    3909.5(4) $&$   13/2^{-} \rightarrow   11/2^{-} $ &		&	&0.97(3)	&	 $M1/E2$ \\
 2345.2(2)     	& 1.5(4)     	&$    x + 5474.3(1) $&$ (J + 6^-) \rightarrow (J + 4^-) $ &			&	&1.14(5)	&	 $E2$ \\
 2445.3(4)       	& 5.8(4)       	&$   14289.7(8) $&$   37/2^{-} \rightarrow   33/2^{-} $ & 0.31(3)	&-0.11(3)	&1.46(7)	&	 $E2$ \\
 2675.7(2)     	& 1.1(3)     	&$    x + 8150.0(9) $&$ (J + 8^-) \rightarrow (J + 6^-) $ &			&	&1.18(4)	&	 $E2$ \\
 2880.2(4)       	& 1.2(5)       	&$   17170.2(3) $&$   (41/2^{-}) \rightarrow   37/2^{-} $ &		&	&	&	 $E2$\footnotemark[1] \\
 2984.2(3)     	& 1.0(1)     	&$   x +  11134.2(8) $&$ (J + 10^-) \rightarrow (J + 8^-) $ &			&	&	&	 $E2$\footnotemark[1]
 \footnotetext[1]{$E2$ multipolarity assumed; see text for details.}
\end{longtable*}
\end{center}
\renewcommand{\thefootnote}{\arabic{footnote}}

\section{Discussion}
The  $^{61}$Co isotope, with $Z=27$ and $N=34$, is located in the upper half of the proton $f_{7/2}$ shell and in the lower part of the $f_{5/2}$, $p_{3/2}$, $p_{1/2}$ neutron subshell. As discussed earlier, this region is susceptible to collective effects at high spins mainly due to the influence of the $\nu g_{9/2}$ intruder orbital which  comes increasingly closer to the Fermi surface with the rise in deformation. At low spins, however, the structure is dominated by single-particle type excitations involving a few nucleons. In this study, the low-spin part will be investigated by comparisons with shell-model predictions. The rotational characteristics of the high-spin bands will be discussed within the framework of a systematic comparison with bands observed in other nuclei in the region. 
\begin{center}
\begin{figure}[b]
\includegraphics[width=0.57\textwidth]{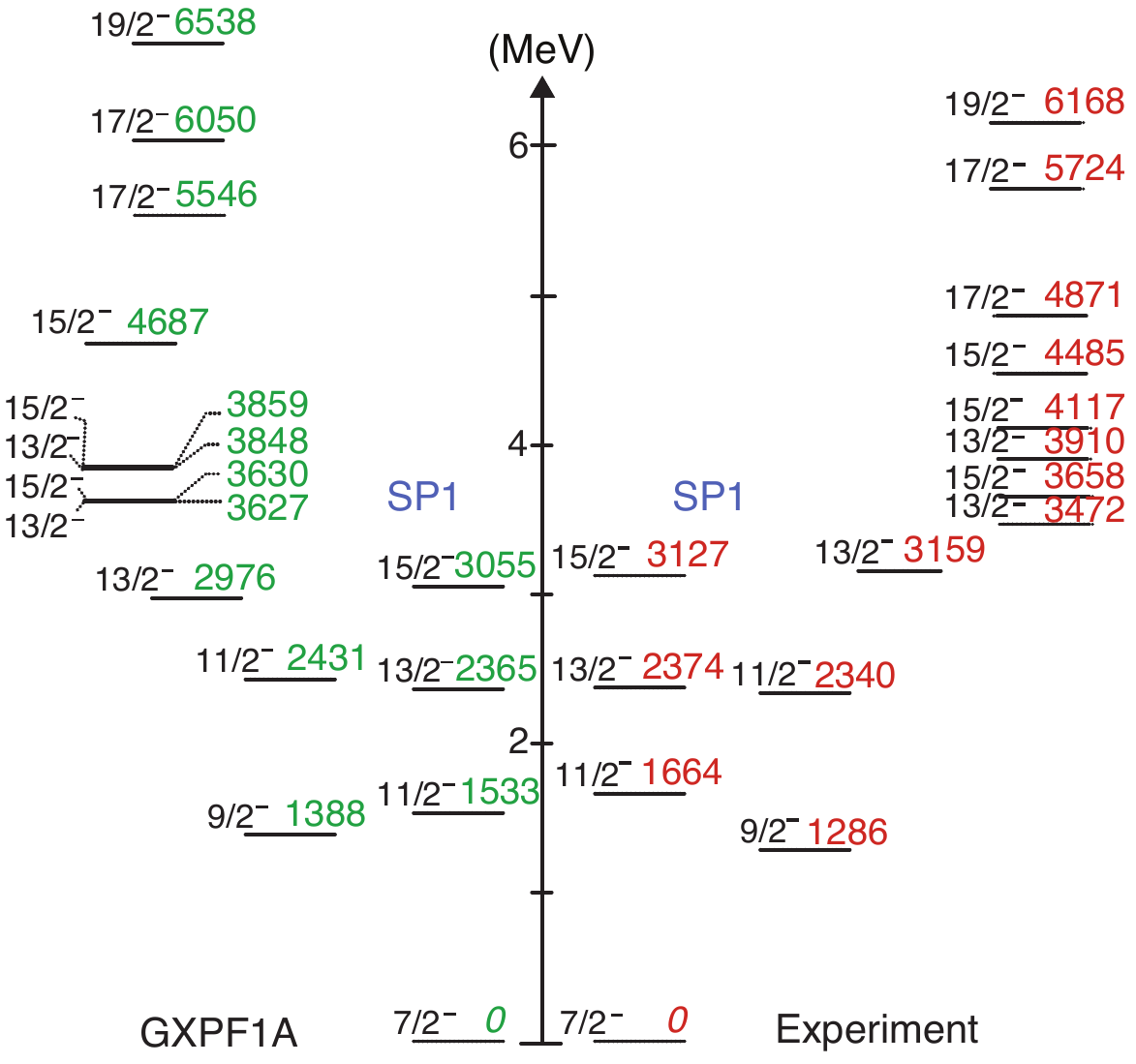}
\caption{\label{fig:shmcalc} (Color online) Shell-model calculations of level structures in $^{61}$Co compared with experimental levels. The picture depicts the single-particle levels marked $SP1$ and the bandheads of the newly identified dipole bands. The calculations used the effective interaction GXPF1A with a $^{40}$Ca closed core.}
\end{figure}
\end{center}

\subsection{Shell-model type excitations}
Shell-model calculations were carried out in the $pf$ model space using the \textsc{antoine} code~\cite{caurier1999,caurier2005}  and the GXPF1A~\cite{Honma2005} effective interaction. With a $^{40}$Ca closed core, the GXPF1A interaction results in a large energy gap between the $\pi f_{7/2}$ and $f_{5/2}$, $p_{3/2}$, and $p_{1/2}$ orbitals such that the proton wave functions for $Z<28$ nuclei are dominated by $\pi(f_{7/2})^n$ configurations. Using this fact, a simple truncation scheme in which the valence protons were confined to the $f_{7/2}$ orbital, and the neutron space restricted to $f_{5/2}$, $p_{3/2}$, and $p_{1/2}$ states, was employed. For the structure under investigation, the truncation scheme appears to work quite well, since the spin and parity of the ground state and the relative energy spacing between levels in the low-spin structure are satisfactorily reproduced. The results of the calculations are compared with the experimental energies in Fig.~\ref{fig:shmcalc}. The calculations include the structure identified as $SP1$ in Fig.~\ref{fig:levelsch}, as well as the bandheads and first excited states in each of the newly identified dipole bands. Following the formalism prescribed in Refs.~\cite{PhysRevC.80.024318,PhysRevC.85.044316-Steppenbeck}, a root-mean-square deviation, $\Delta_{\mathrm{rms}}$, between the calculated and experimental energies was used to measure the degree of agreement. For states below 3.5 MeV and $I^\pi \le 15/2^-$, a $\Delta_{\mathrm{rms}}$ value of 120 keV, corresponding to an average energy difference between states of less than 150 keV, was achieved. This suggests that these states are mostly characterized by single-particle excitations. Above 3.5 MeV, significant rms deviations from the experimental energies are observed. For example, the calculations result in energy separations far less than the experimental values: between the $J^\pi=13/2^-$ and $15/2^-$ levels; i.e., the bandhead and first excited state in band $DB2$, the calculated separation is only 3 keV, while the computed difference between the first excited level and the bandhead in band $DB3$ is only 11 keV. Furthermore, the calculations predict excitation energies that are substantially higher than the experimental counterparts for states beyond the $17/2^-$ level. These discrepancies indicate that, at higher excitation energies and spins, the model space used in the calculations becomes inadequate for a proper description of the levels. It also implies that these states (the bandheads), and the bands built on top of them, are likely associated with a degree of deformation. In fact, the increasing collectivity reflected in experimental data for excitation energies above 4 MeV could not be reproduced in the calculations. This is not unexpected since the $fp$ shell model space of the GXFP1A interaction does not include the $g_{9/2}$ orbital needed for a successful description of the spectrum of higher-lying states. 

\begin{center}
\begin{figure}[t!]
\includegraphics[width=0.47\textwidth]{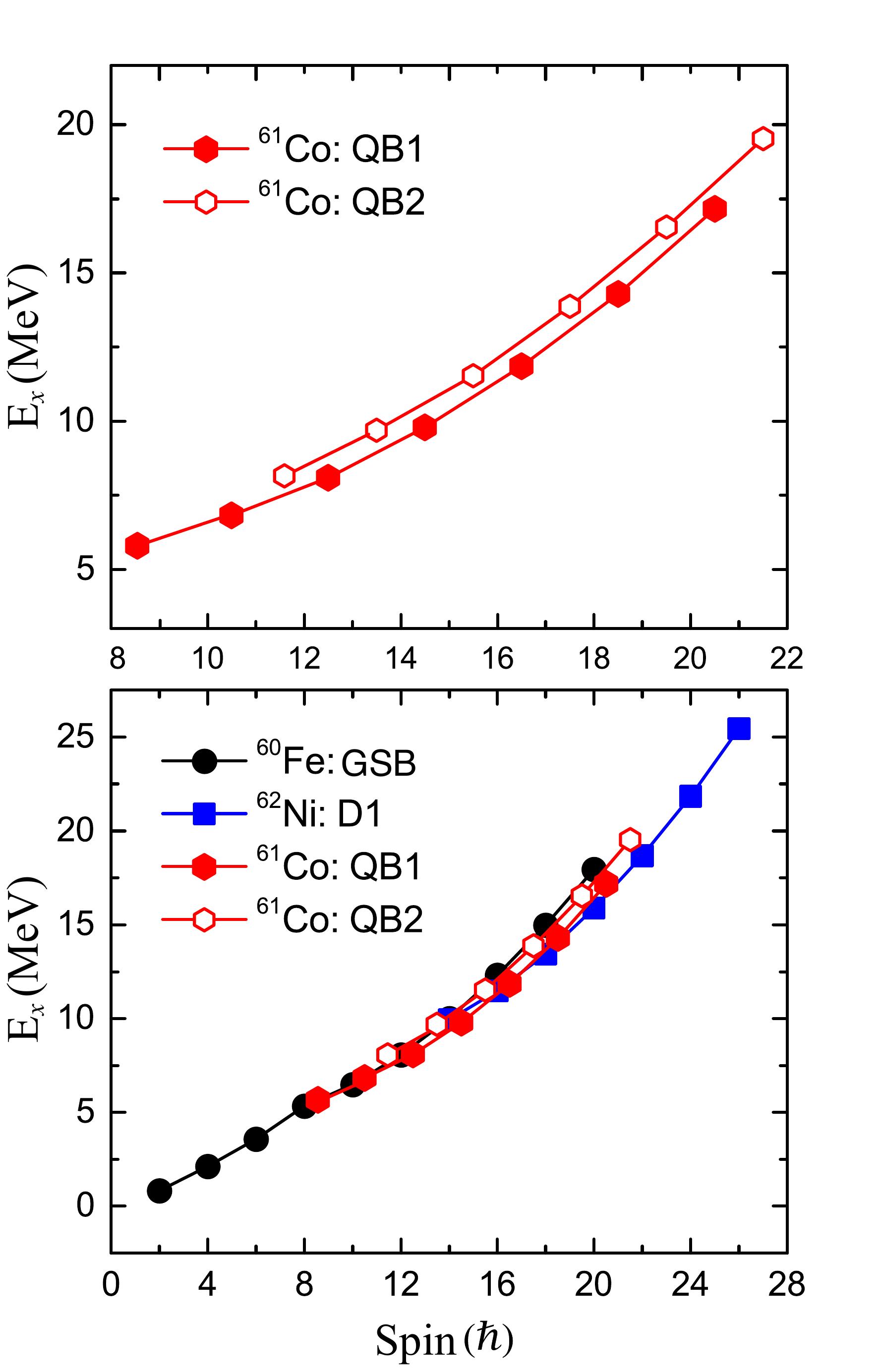} 
\caption{\label{fig:exciten} (Color online) Excitation energy $E_x$ as a function of spin for the $\Delta I=2$ bands in $^{61}$Co observed in the present investigation. The top panel shows the bands alone, while the lower one compares the new bands with similar sequences in the $N=34$ isotones $^{60}$Fe and $^{62}$Ni. Data and band names for the isotones are taken from Refs.~\cite{PhysRevC.76.054303} and \cite{ni62-albers}. Note that different scales are used in the top and bottom panels.}
\end{figure}
\end{center}

\begin{center}
\begin{figure}[t!]
\includegraphics[width=0.47\textwidth]{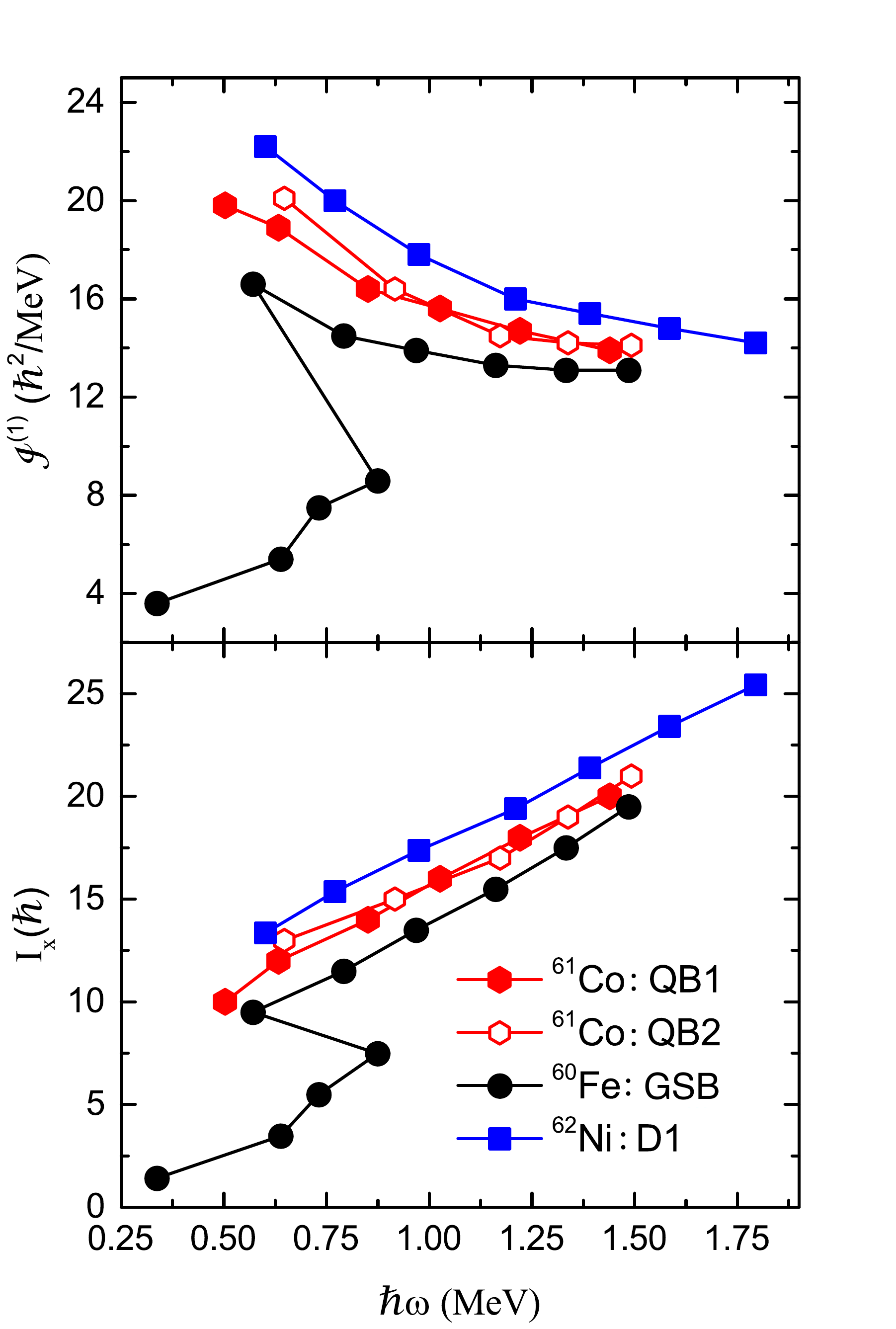} 
\caption{\label{fig:ixplot} (Color online) (Top) Experimental kinematic moment of inertia $\mathcal{J}^{(1)}$ as a function of the rotational frequency $\hbar \omega$ for bands $QB1$ and $QB2$ in $^{61}$Co, the yrast band in $^{60}$Fe ($GSB$ in Ref.~\cite{PhysRevC.76.054303}), and band $D1$ in $^{62}$Ni~\cite{ni62-albers}. (Bottom)  Experimental alignment $I_x$  as a function of the rotational frequency $\hbar \omega$ for bands $QB1$ and $QB2$ in $^{61}$Co, the yrast band in $^{60}$Fe ($GSB$ in Ref~\cite{PhysRevC.76.054303}) and band $D1$ in $^{62}$Ni~\cite{ni62-albers}. Data and band names for the isotones are taken from Refs.~\cite{PhysRevC.76.054303} and \cite{ni62-albers}.}
\end{figure}
\end{center}

\vspace{-2.7cm}
\subsection{Collective excitations}
\subsubsection{\label{qbands}Quadrupole Bands}
In light of the above discussion, it appears that any description of the higher-lying states in $^{61}$Co will have to be carried out in an expanded model space beyond the $pf$ shell. Several  $\Delta I =2 $ bands have been observed to high spins in nuclei of the $A\sim60$ mass region which have been interpreted by invoking configurations based on the $g_{9/2}$ orbital. For example, the levels for $I > 6$ of the yrast sequence in $^{60}$Fe are described in terms of a rotational band based on a $\nu (g_{9/2})^2$ configuration~\cite{PhysRevC.76.054303}. This sequence is associated with an axially deformed nuclear shape, with a characteristic deformation parameter of $\beta_2 \sim 0.2$. Similarly, rotational band structures with fairly large deformation have also been observed in  $^{62}$Ni~\cite{ni62-albers} and $^{63}$Ni~\cite{PhysRevC.88.054314-albers}. Figure~\ref{fig:exciten}(a) provides the excitation energies, $E_x$, as a function of spin for the two $\Delta I  =2$ bands, $QB1$ and $QB2$, in $^{61}$Co obtained in the present investigation. For band $QB2$, the limit values for excitation energies and spin discussed above have been adopted. These data are compared with the high-spin yrast sequences in the $N=34$ isotones, $^{60}$Fe and $^{62}$Ni, in Fig.~\ref{fig:exciten}(b). The trajectories in the ($E_x, I)$ plane for the two bands are close to that of the yrast sequence in $^{60}$Fe and also follow the same trajectory as $^{62}$Ni ($D1$ in Ref.~\cite{ni62-albers}), strongly suggesting through their similar pattern that one is dealing with collective excitations of the same character. Figure~\ref{fig:ixplot}(a) provides the kinematic moments of inertia of bands $QB1$ and $QB2$ compared with the yrast high-spin bands in $^{60}$Fe and $^{62}$Ni. The aligned angular momenta $I_{\mathrm{x}}$ as a function of the rotational frequency $\omega$ for the two bands in $^{61}$Co in comparison with the sequences in the $N=34$ isotones are presented in Fig.~\ref{fig:ixplot}(b). It can be seen from both figures that the bands all exhibit similar dynamical behavior at high spins. The sudden change in the $I_x$ trajectory for the yrast band in $^{60}$Fe (black filled circles) is attributed to the crossing, at a rotational frequency of $\hbar \omega \approx 0.5$ MeV, of the ground-state sequence with a rotationally-aligned band built on the $\nu (g_{9/2})^2$ configuration. Above the crossing frequency, the extracted aligned angular momenta for band $QB1$ in $^{61}$Co (red filled hexagons) follow a similar trajectory as a function of frequency and exhibit the same gradient as the $\nu (g_{9/2})^2$ band in $^{60}$Fe. This suggests that the configuration associated with band $QB1$ also includes $\nu (g_{9/2})^2$ rotationally-aligned neutrons, with a small additional gain in alignment ($\sim$$2\hbar$), (i.e., the gap between the $I_x$ trajectories of $^{60}$Fe and $^{61}$Co) probably associated with a contribution by the extra proton of $^{61}$Co. Likewise, the $I_x$ curve for $^{62}$Ni (blue filled squares) exhibits the same $\nu (g_{9/2})^2$ behavior above the crossing frequency, indicating that it might be associated with a four-quasiparticle configuration, with the resultant gain in alignment ($\sim$$4\hbar$) being provided by an additional two protons or two neutrons (quasiparticles) relative to a $^{60}$Fe core. In addition to having the same rotational characteristics, the extracted deformation parameters for the high-spin yrast bands in $^{61}$Co and $^{62}$Ni are of the same magnitude ($\beta \sim 0.2 - 0.4$) within admittedly large errors, and in the same range as those reported in $^{63}$Ni. The $I_x$ values for band $QB2$ (red open hexagons) were again extracted based on the assumptions defined above and adopted in Fig.~\ref{fig:exciten}; i.e., with a bandhead energy $\ge$ 8.4 MeV and a spin-parity value $\ge$ $23/2^-$. With reference to the $\nu (g_{9/2})^2$ trajectories, the $I_x$ trajectory for this band follows a rather similar path, indicating that this structure likely involves a $\nu (g_{9/2})^2$ configuration as well. %
\begin{center}
\begin{figure}[t!]
\includegraphics[width=0.47\textwidth]{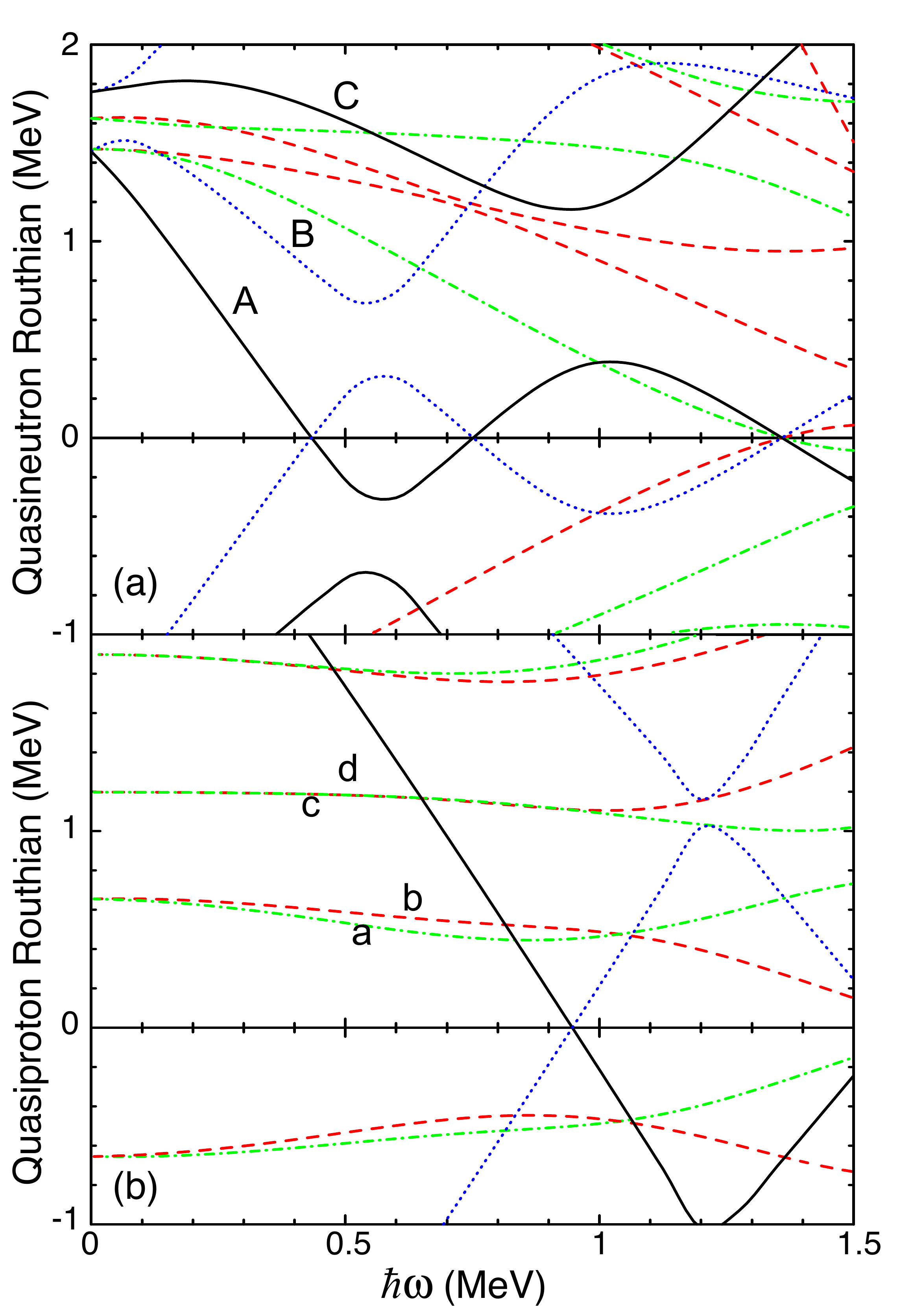} 
\caption{\label{fig:csmnp3} (Color online) Results of cranked shell-model calculations (CSM) for the quasineutron (a) and quasiproton (b) routhians as a function of rotational frequency $\hbar \omega$ in $^{61}$Co. The calculation was performed with parameters of $\beta_2 = 0.3$, $\beta_4=0.0$, and $\gamma =0^\circ$. The line types represent unique combinations of (parity, signature), as follows: Solid $(+,+1/2)$; Dot $(+,-1/2)$; Dash-dot $(-,+1/2)$; Dash $(-,-1/2)$. Quasiparticle labels follow the convention of Ref.~\cite{Bengtsson198615}. See text for details.}
\end{figure}
\end{center}

In order to provide a qualitative and microscopic description of the variations in the alignment properties of the observed sequences with rotational frequency within the framework of the cranked shell model (CSM), calculations were performed for both quasiprotons and quasineutrons in $^{61}$Co. 
The level energies were calculated in a deformed Wood-Saxon potential with universal parameters using a deformation parameter set ($\beta_2 = 0.3$, $\beta_4 = 0.0$, and $\gamma = 0^\circ$) chosen to be within the limits of error of the experimentally observed value (see above). The pairing energies at zero frequency $\Delta_\mathrm{n} (\omega=0) = 1.4542$ MeV and $\Delta_\mathrm{p} (\omega=0) = 0.6473$ MeV were determined using the BCS formalism and kept constant as a function of frequency. The resulting  quasiproton and quasineutron routhians are presented in Figs.~\ref{fig:csmnp3}(a) and (b), respectively, and the relevant quasiparticle orbitals (i.e., those near the Fermi surface) are labeled using the convention of Ref.~\cite{Bengtsson198615}. These are summarized in Table~\ref{tab:csm}.

\begin{table}[h!]
\caption{\label{tab:csm}Description of the labeling convention for the quasiparticle orbitals close to the Fermi surface in the CSM calculations for $^{61}$Co.}
\begin{ruledtabular}
\begin{tabular}{lcc}
Shell & Nilsson Label & ($\alpha = +\frac{1}{2},-\frac{1}{2}$) \\
\hline
$\nu \mathrm{g}_{9/2}$ & $[440]1/2$ & A, B \\
$\nu \mathrm{g}_{9/2}$ & $[431]3/2$ & C, D \\
$\pi \mathrm{p}_{3/2}$ & $[321]1/2$ &  a, b\\
$\pi \mathrm{f}_{7/2}$ & $[303]7/2$ &  c, d\\
\end{tabular}
\end{ruledtabular}
\end{table}

As shown in Fig.~\ref{fig:csmnp3}(a), the lowest-lying quasineutron routhians are associated with the $g_{9/2}$ orbital, favoring a $\nu (g_{9/2})^2$ configuration in agreement with the analysis presented earlier. The theoretically predicted crossing ($AB$) at a frequency of $\hbar \omega \sim 0.5$ MeV is not observed experimentally in the two bands in $^{61}$Co due to the fact that the $A$ and $B$ orbitals are already occupied for these bands.  It should be noted that this crossing is different from the experimental backbend at a similar frequency [see Fig.~\ref{fig:ixplot}(b)] for $^{60}$Fe, which has been interpreted as resulting from the interaction between the low-spin shell-model states and a collective band built on a $\nu(g_{9/2})^2$ aligned neutron configuration~\cite{PRC.87.041305-mpcarpenter}. For the quasiproton routhians [Fig.~\ref{fig:csmnp3}(b)], the lowest-lying quasiparticle orbitals are associated with the $[321]\frac{1}{2}$ Nilsson orbit of $p_{3/2}$ parentage, which is lowered in energy relative to the orbitals of $f_{7/2}$ parentage due to deformation, and exhibits a small but distinguishable signature splitting. Comparing this observation to Fig.~\ref{fig:ixplot}(b) immediately reveals that bands $QB1$ and $QB2$ might be signature partners built on the same $p_{3/2}$ proton configuration coupled to the $\nu(g_{9/2})^2$ configuration. Thus, a consistent picture appears to emerge. However, the assignment for band $QB2$ remains tentative since the relative excitation energies of the two bands have not been established. Further work will be required in order to either validate or modify this interpretation. 

\subsubsection{Dipole Bands}
In addition to the two quadrupole sequences discussed above, four $\Delta I=1$ bands ($DB1 - 4$) without $E2$ cross-over transitions, but exhibiting rather regular patterns, were also delineated to fairly high spins. As discussed in Section~\ref{levelsch} above, these bands have been linked to the single-particle structure, thus enabling firm determination of spins, parities, and excitation energies. In general, the occurrence of such $\Delta I=1$ bands is often associated with one-particle one-hole excitations involving high-$K$ proton holes. In this $^{61}$Co case, such an interpretation would have to involve the $f_{7/2}$ proton hole and would appear to be rather unlikely as candidate configurations would require fairly large deformations that, in turn, would favor the presence of $E2$ radiation competing with the dipole strength in the decay of the states. Furthermore, the lack of signature splitting and departures of the excitation energies from the conventional rotational behavior (see below) make such an interpretation doubtful. On the other hand, as will be shown below, the dipole bands share distinct characteristics with two $\Delta I=1$ bands firmly established recently in $^{58}$Fe~\cite{PhysRevC.85.044316-Steppenbeck}, where an interpretation in terms of the shears mechanism (i.e., magnetic rotation) was proposed, based on calculations within the self-consistent tilted axis cranking relativistic mean field model (TAC-RMF)~\cite{PhysRevC.78.024313,Zhao2011181,PhysRevC.85.044316-Steppenbeck}. Furthermore, $\Delta I=1$ bands have also been tentatively reported in $^{60}$Ni~\cite{PhysRevC.78.054318}, and a possible interpretation in terms of the shears mechanism has been proposed in this instance as well. Hence, the possibility that the $DB1 - 4$ bands are associated with the same mechanism deserves closer examination.

According to Ref.~\cite{PhysRevC.85.044316-Steppenbeck}, in the $A\sim 60$ mass region, magnetic rotation originates from the alignment of angular momentum vectors built on high-$\Omega$ proton holes associated with the $f_{7/2}$ orbital, coupled to $j_\pi$, and low-$\Omega$ $g_{9/2}$ neutron states, coupled to $j_\nu$. At the bandhead, the two angular momentum vectors ($\vec{j}_\pi$  and  $\vec{j}_\nu$) are approximately perpendicular to one another and the angular momenta of higher-energy states along a band increase by the gradual recoupling of the two spin vectors in the direction of the total angular momentum in a manner reminiscent of the progressive closing of the blades of a shears~\cite{Frauendorf1993259}. Following the review of Ref.~\cite{annurev.nucl.50.1.1}, this recoupling results in excitation energies of states within a dipole band with rotational-like behavior described by the expression: $E_x(I) -E_0 \sim A(I - I_0)^2$ where $I$ is the spin of a level of interest and $I_0$ is the spin of the bandhead. 

\begin{center}
\begin{figure}[t!]
\includegraphics[width=0.7\textwidth]{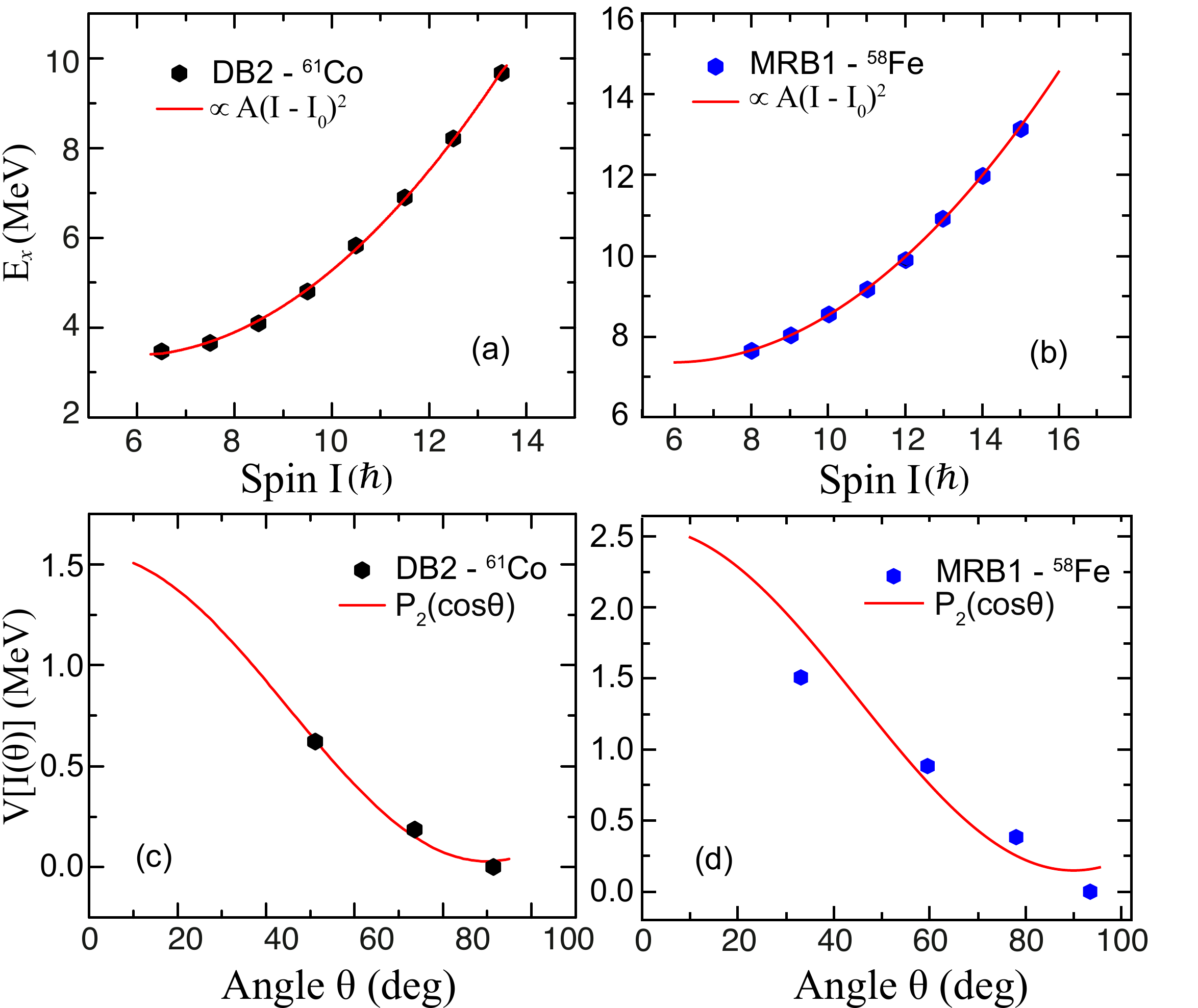} 
\caption{\label{fig:dbands} (Color online) Excitation energies $E_x$ of states as a function of spin $I$ for (a) band $DB2$ in $^{61}$Co and (b) band $MRB1$ in $^{58}$Fe. Effective interaction as a function of the shears angle $\theta$ for (c) band $DB2$ in $^{61}$Co and  (d) band $MRB1$ in $^{58}$Fe. The solid lines in panels (c) and (d) show the dependence of a $P_2$-type force.}
\end{figure}
\end{center}

Focusing first on the yrast dipole band, $DB2$, Fig.~\ref{fig:dbands}(a) provides the excitation energy as a function of angular momentum for this sequence and compares it with that of band $MRB1$ of $^{58}$Fe~\cite{PhysRevC.85.044316-Steppenbeck} in Fig.~\ref{fig:dbands}(b). The solid curves in both panels represent fits with the expression above. The behavior of both bands is strikingly similar. In particular, unlike usual rotational bands, both curves display a minimum in excitation energy at a non-zero angular momentum, $I_o$, indicating that the behavior does likely not originate from quadrupole collective motion. The noted similarity between the $^{61}$Co and $^{58}$Fe bands can be viewed as a first argument in favor of the shears band interpretation. To investigate the properties of bands $DB2$ and $MRB1$ further in the framework of the shears mechanism, the semi-classical approach of Refs.~\cite{AOM-PRC.57.R1073,AOM-PRC.58.R621} was employed. This schematic model interprets the rotational-like behavior of the $\Delta I = 1$ bands as a consequence of the residual interaction between the proton and neutron blades of the shears and introduces the shears angle $\theta$ between the two angular momentum vectors $\vec{j}_\pi$ and  $\vec{j}_\nu$ as a degree of freedom defined by the expression $\mathrm{cos}\:\theta=(I^2-j_\pi^2 - j_\nu^2)/2 j_\pi j_\nu$. In the case of band DB2, perpendicular coupling at the bandhead results in the values $j_\pi = 7/2$ and $j_\nu$ $\approx 11/2$ in order to reproduce the $I_0 = 13/2$ spin.  

Pursuing further the procedure outlined in Refs.~\cite{AOM-PRC.57.R1073,AOM-PRC.58.R621}, the total angular momentum along the band, decomposed into $I =I_{\mathrm{shears}} + R_{\mathrm{core}}$, points to a core contribution of less than $10\%$ at the top of the DB2 band. Hence, in this approach, over $90\%$ of the angular momentum along the sequence can be assigned to the shears mechanism and, consequently, the energy required to generate the shears at each spin originates from the change in potential energy generated by the recoupling of the angular momenta of the proton and neutron blades. As demonstrated in Ref.~\cite{AOM-PRC.57.R1073}, the latter energy is given by: $V[I(\theta)] = E_x(I) - E_0$. A correlation between the effective interaction, defined by the change in potential generated by the blades, and the angle between them can be deduced: the resulting smooth variation of $V$ with $\theta$  is presented in panels (c) and (d) of Fig.~\ref{fig:dbands} for bands $DB2$ and $MRB1$, respectively. When allowing contributions from spatial forces only, this effective interaction can then be expanded~\cite{annurev.nucl.50.1.1} in terms of even multipoles as $V(\theta)=V_0\;  + \; V_2 P_2 (\theta) \; +  \; \dotsb$. The strength of the $V_2$ interaction extracted from Fig.~\ref{fig:dbands}(c) is positive, as anticipated for a particle-hole coupling, and of the order of $\sim$$1$ MeV, a value which compares well with the $\sim$$1.3$ MeV interaction derived in a similar analysis of the $MRB1$ band in $^{58}$Fe. It should be noted that the strength distributions derived for these two nuclei do not strictly follow the $1/A$ scaling observed in, for example, the well-documented shears bands of the Pb isotopes~\cite{PhysRevLett.78.1868,PhysRevC.58.R1876}, although similar deviations have been reported in the $A\sim110$ region as well~\cite{Clark-PRLett.82.3220,PhysRevC.61.034318}. Reference~\cite{annurev.nucl.50.1.1} suggests the presence of successive shears along a dipole sequence as a possible explanation for such deviations, but such an interpretation is beyond the scope of the present work. 

\vspace{-1.4cm}
The rudimentary analysis presented here reinforces the striking similarity between the observations in $^{61}$Co and $^{58}$Fe and naturally leads one to adopt a similar interpretation in terms of magnetic rotation. However, placing this interpretation on stronger footing requires further work. On the experimental side, a measurement of the electromagnetic transition probabilities along the sequence would be necessary. On the theory side, calculations along the lines of those presented in Ref.~\cite{PhysRevC.85.044316-Steppenbeck} are desirable. Work in this direction is underway, but has yet to account for the present results~\cite{pwzhao}. It is worth noting that the two candidate configurations proposed for the $MRB1$ band in $^{58}$Fe following the TAC-RMF calculations of Ref.~\cite{PhysRevC.85.044316-Steppenbeck} involve both $f_{7/2}$ proton holes and $g_{9/2}$ neutrons and one would expect these states to be involved here as well.    

\begin{center}
\begin{figure}[]
\includegraphics[width=0.7\textwidth]{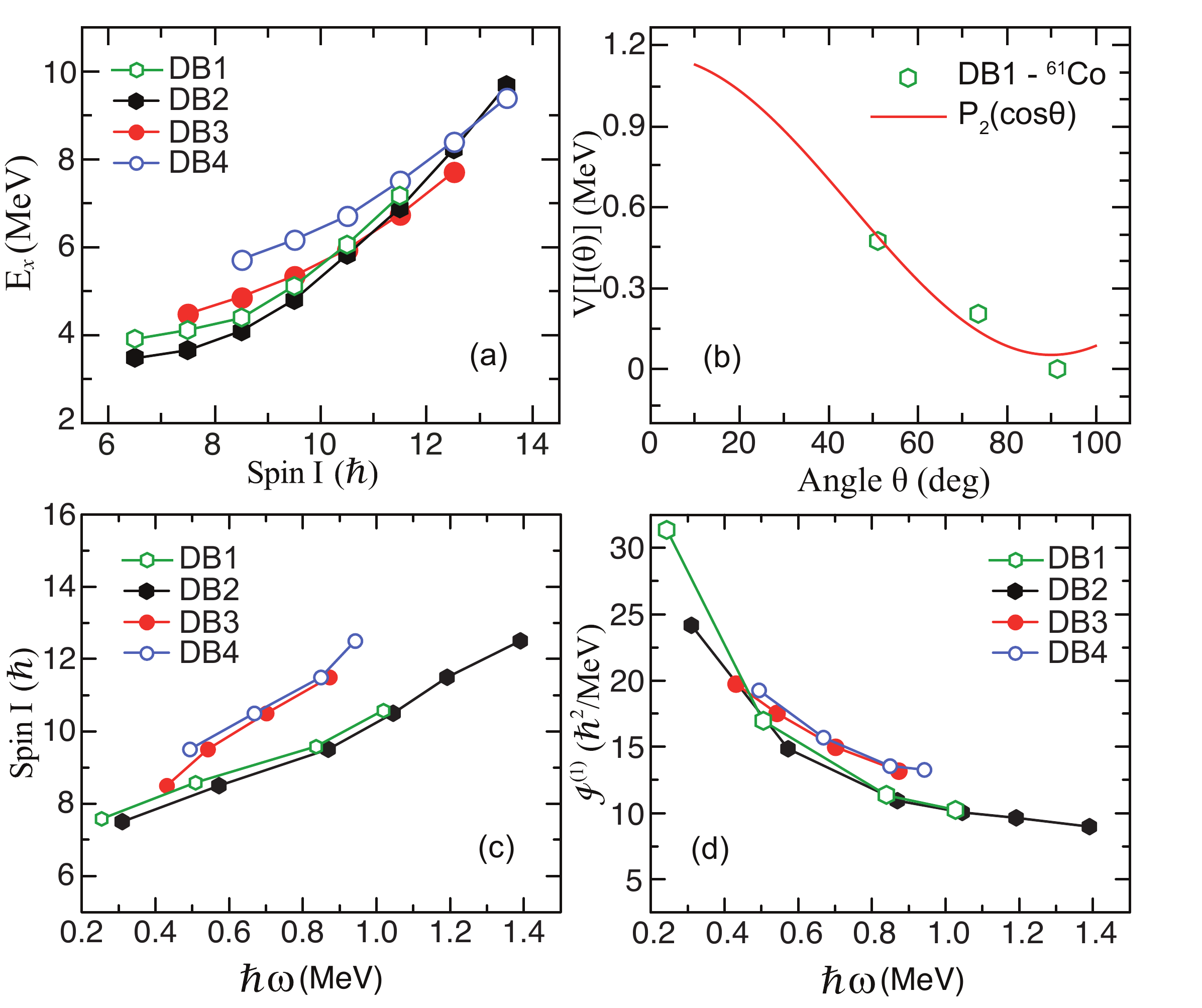} 
\caption{\label{fig11} (color online) (a) Excitation energy as a function of spin for all the observed $\Delta I =1$ bands in $^{61}$Co. (b) Effective interaction as a function of the shears angle for band $DB1$ in $^{61}$Co. The solid line shows the dependence of a $P_2$-type force.  (c) Angular momentum and (d) kinematic moment of inertia $\mathcal{J}^{(1)}$ versus rotational frequency $\hbar \omega$ for bands  $DB1$, $DB2$, $DB3$ and $DB4$.}
\end{figure}
\end{center}

The various panels of Fig.~\ref{fig11} provide comparisons of the properties of the three additional dipole bands ($DB1$, $DB3$, and $DB4$) in $^{61}$Co, with those of sequence $DB2$. Particularly striking in panel (a) is the similarity of the trajectories of bands $DB1$ and $DB2$ in the $(E_x, I)$ plane: the two sequences mirror one another up to $I\sim{23/2}$, the highest spin in the former. As a result, an analysis in terms of the shears mechanism, similar to that introduced above, was also carried out in the case of band $DB1$. The calculation of the interaction strength as a function of the angle between the blades of panel (b) leads to $V_2$ value of $\sim$$1.1$ MeV, in line with the strengths discussed above. Thus, just like in the case of $^{58}$Fe, at least two bands in $^{61}$Co can be viewed as candidates for an interpretation in terms of magnetic rotation. Panels (c) and (d) of Fig.~\ref{fig11} extend the comparison to the total angular momentum and kinematic moments of inertia $\mathcal{J}^{(1)}$, both as a function of the rotational frequency. Panel (c) highlights the nearly linear increase of $I$, and panel (d) the monotonic decrease of $\mathcal{J}^{(1)}$, with $\hbar \omega$. Both of these observations are additional characteristics associated with magnetic rotation~\cite{annurev.nucl.50.1.1}, herewith reinforcing the suggested interpretation. Figure~\ref{fig11} also points to similarities between the properties of bands $DB3$ and $DB4$: again, in every panel, the trajectories of the bands are rather similar, while somewhat distinct from those for bands $DB1$ and $DB2$. In fact, these two additional sequences appear to be nearly degenerate in the figure. Unfortunately, the data on bands $DB3$ and $DB4$ are limited. As a result, an interpretation invoking some of the phenomena sometimes associated with near degeneracies such as either chiral doubling~\cite{frauendorf2001spontaneous} or the presence of pseudospin doublet bands~\cite{Arima1969517,Hecht1969129}, is premature at this time.

\section{Conclusions}
The structure of the odd-mass $^{61}$Co nucleus has been considerably expanded by using a complex, multi-nucleon transfer reaction in inverse kinematics, exploiting the sensitivity of an experimental setup combining Gammasphere with the Fragment Mass Analyzer (FMA). Shell-model calculations, carried out with the GXPF1A effective interaction in a modest-size $pf$ space, successfully describe the low-spin structure, confirming the view that the levels in this region are mostly associated with particle-hole excitations. However, at higher spins, strong evidence for collective behavior was uncovered. Two quasi-rotational bands of stretched-$E2$ transitions were established up to spins of $I\sim 41/2$ and excitation energies $> 17$ MeV. Based on the measured Doppler shifts, the bands were determined to be associated with a sizable $\beta_2$ deformation. The two bands can be understood by combining comparisons with rotational bands in neighboring $A\sim 60$ nuclei and results of cranked shell-model calculations as rotational sequences built on configurations involving two $g_{9/2}$ neutrons coupled to the two signatures of the $p_{3/2}$ proton orbital. Furthermore, four dipole bands were observed, and were traced over a relatively wide spin range. Based on comparisons with observations in the neighboring $^{58}$Fe nucleus, and aided by an analysis based on a semi-classical description of the coupling of angular momenta within the shears mechanism, two of these sequences are proposed to be associated with magnetic rotation of a nearly spherical nucleus.  The picture that emerges from the present work reinforces the view that, once $g_{9/2}$ neutron excitations become energetically favored, collectivity occurs and the motion can be associated with different nuclear shapes. Thus, despite the limited number of orbitals present near the Fermi surface, nuclei of the $A\sim60$ region display a rich variety of phenomena similar to those seen in heavier systems.  In $^{61}$Co, quadrupole collectivity associated with a prolate shape competes for yrast status with magnetic rotation of a nearly spherical system.

\section{Acknowledgments}
This material is based upon work supported by the U.S. Department of Energy, Office of Science, Office of Nuclear Physics under Contract number DE-AC02-06CH11357, and under Award Number DE-FG02-94ER40834 and DE-FG02-08ER41556, and by  the National Science Foundation under Contract PHY-1102511, and by the United Kingdom Science and Technology Facilities Council (STFC). This research used resources of ANLÕs ATLAS facility, which is a DOE Office of Science User Facility.

\bibliographystyle{apsrev4-1}

\bibliography{Co61}

\end{document}